\documentclass[a4paper,fleqn,usenatbib]{mnras}

\usepackage{newtxtext,newtxmath}

\usepackage[T1]{fontenc}
\usepackage{ae,aecompl}

\usepackage{graphicx}	
\usepackage{amsmath}	
\usepackage{amssymb}	

\title[Galaxy size versus stellar mass relation]{Analysis of the galaxy size versus stellar mass relation}

\author[\,S\'anchez~Almeida]{
  J.\,S\'anchez~Almeida$^{1,2}$\thanks{E-mail: jos@iac.es}\\
$^1$Instituto de Astrof\'\i sica de Canarias, La Laguna, Tenerife, Spain\\
$^2$Departamento de Astrof\'\i sica, Universidad de La Laguna, Tenerife, Spain
}

\date{Accepted XXX. Received YYY; in original form Feb 2020}

\pubyear{2020}

\begin{document}
\label{firstpage}
\pagerange{\pageref{firstpage}--\pageref{lastpage}}
\maketitle

\begin{abstract}
  The scatter in the galaxy size versus stellar mass ($M_\star$) relation gets largely reduced when, rather than the half-mass radius $R_e$, the size at a fixed surface density is used. Here we address why this happens.  We show how a reduction is to be expected because any two galaxies with the same $M_\star$ have at least one radius with identical surface density, where the galaxies have identical size.  However, the reason why the scatter is reduced to the observed level is not trivial, and we pin it down to the galaxy surface density profiles approximately following Sersic profiles with their $R_e$ and Sersic index ($n$) anti-correlated (i.e., given $M_\star$, $n$ increases when $R_e$ decreases). Our analytical results describe very well the behavior of the observed galaxies as portrayed in the NASA Sloan Atlas (NSA), which contains more than half a million local objects with $7 < \log(M_\star/M_\odot) < 11.5$. The comparison with NSA galaxies also allows us to find the optimal values for the mass surface density ($2.4_{-0.9}^{+1.3}\,M_\odot\,{\rm pc}^{-2}$) and surface brightness ($r$-band  $24.7\pm 0.5\,{\rm mag\,arcsec^{-2}}$) that minimize the scatter, although the actual values depend somehow on the subset of NSA galaxies used for optimization. The physical reason for the existence of optimal values is unknown but, as \citet{2020arXiv200102689T} point out, they are close to the gas surface density threshold to form stars and thus may trace the physical end of a galaxy. Our NSA-based size--mass relation agrees with theirs on the slope as well as on the magnitude of the scatter.
As a by-product of the narrowness of the size--mass relation (only 0.06 dex), we propose to use the size of a galaxy to measure its stellar mass. In terms of observing time, it is not more demanding than the usual photometric techniques and may present practical advantages in particular cases.
\end{abstract}




\section{Introduction} \label{sec:intro}

The proper way to measure galaxy size has been discussed in the literature for long. Astronomers have used radii containing a fix fraction of the integrated galaxy light \citep[e.g.,][]{1948AnAp...11..247D, 1980ApJS...43..305K, 1976ApJ...209L...1P} and also radii up to a given surface brightness \citep[e.g.,][]{1958MeLuS.136....1H,1960ApJ...132..306L,1976srcb.book.....D,1975A&A....40..133P}.
The extremely influential Reference Catalogues of Bright Galaxies  \citep[RC2 and RC3,][]{1976srcb.book.....D,1991rc3..book.....D}  parameterize galaxy size as the major axis diameter at a surface brightness of 25 mag~arcsec$^{-2}$. Other examples of systematic compilations of sizes using isophotal radii are
the Uppsala General Catalogue of galaxies \citep[UGC,][]{1973ugcg.book.....N},
the Lyon-Meudon Extragalactic Database \citep[LEDA,][]{1989A&AS...80..299P,1996A&A...311...12P},
the ESO/Uppsala survey of the southern sky \citep[][]{1974A&AS...18..463H,1975A&AS...22..327H},
several common nearby galaxy catalogs \citep{2004AJ....127.2031K,2013AJ....145..101K,1992AN....313..189S,1998ARA&A..36..435M},
and the celebrated HI catalogue by \citet[][]{1981ApJS...47..139F}.
With the substitution of photographic plates with CCD detectors, 
the use of radius containing a fixed fraction of the light (typically 50\,\% or 90\,\%) gradually replaced size determinations based on surface brightness limits. These galaxy sizes proved to be more robust against noisy images \citep[e.g.,][]{2001MNRAS.326..869T}, are conveniently provided by the used automatic analysis techniques
(e.g., SExtractor, \citeauthor{1996A&AS..117..393B}~\citeyear{1996A&AS..117..393B} or  GALFIT, \citeauthor{2002AJ....124..266P}~\citeyear{2002AJ....124..266P}),
and are relatively straightforward to compare with predictions from numerical simulations (e.g., EAGLE, \citeauthor{2015MNRAS.446..521S}~\citeyear{2015MNRAS.446..521S} or Illustris, \citeauthor{2014MNRAS.444.1518V}~\citeyear{2014MNRAS.444.1518V}). 
They are now included in the most popular galaxy surveys,  e.g.,
SDSS \citep[][]{2002AJ....123..485S},
2MASS \citep[][]{2003AJ....125..525J}, 
%
CANDELS \citep[][]{2011ApJS..197...35G}, or 
%
DES \citep[][]{2018ApJS..239...18A},
%
%
even though the current surveys often include isophotal radii too.

Recently,   \citet{2020arXiv200102689T} propose using a size set by the  gas surface density threshold needed for the gas to form stars. In practice, their approach crystalized in $R_1$ defined as  the radius at which galaxies reach a {\em stellar mass surface density} of $1\,M_\odot\,{\rm pc}^{-2}$. $R_1$ has several conceptual and technical advantages with respect to previous definitions: (1) it is physically motivated since this surface density traces to the gas density threshold to form stars, (2) it agrees with the radius where the truncation of the disk occurs, (3) it coincides with the physical end of the disk as inferred from state-of-the-art deep optical images, and (4) the scatter of the relation size versus stellar mass ($M_\star$) is largely reduced compared to the scatter when the effective radius (i.e., the half-light radius $R_e$) or other definition of size is employed.
The present paper is aimed at analyzing and shedding light on the 4th item, namely, on why the scatter of the size versus $M_\star$ relation is reduced. In answering this question,  we also have to address the issues of whether $1\,M_\odot\,{\rm pc}^{-2}$ is optimal, why disk truncations approximately happens at $R_1$, and what does it imply in terms of galaxy physics.

A significant part of the argumentation in the paper is based on using Sersic functions to represent surface brightness profiles of galaxies. They were introduced by  \citeauthor{1963BAAA....6...41S}~(\citeyear{1963BAAA....6...41S},\citeyear{1968adga.book.....S}) to describe early type galaxies, but they also account the exponential drop characteristic of the late type galaxies. Sersic profiles were barely used for 20 years until the early 90's when it became clear that they represented the observed light distribution significantly better than the traditional {\em de Vaucouleurs $R^{1/4}$ law} (\citeauthor{1993MNRAS.265.1013C}~\citeyear{1993MNRAS.265.1013C},  \citeauthor{1994MNRAS.271..523D}~\citeyear{1994MNRAS.271..523D}; see  \citeauthor{2005PASA...22..118G}~\citeyear{2005PASA...22..118G}  for an historical account). Nowadays, they are routinely used in all large surveys and provided by the main photometric analysis tools (see the above references).  However, it is clear that they only offer a schematic representation since, when the angular resolution and sensitivity of the images are good enough, departures from the single Sersic profile behavior appear,  and several components are needed to model galaxies from center to outskirt \citep[e.g.,][]{1995MNRAS.275..874A,1998MNRAS.299..672S,1999ASPC..176..402I,2001A&A...367..405P}.
The breakdown also occurs at very low surface brightness or low signal to noise ratio \citep[e.g.,][]{2015ApJS..219....4S}, like at high redshift \citep[e.g.,][]{2012ApJS..203...24V}, and when significant dust attenuation come into play \citep[e.g.,][]{2008MNRAS.388.1708G}.
Even with all these caveats in mind, Sersic functions provide a schematic yet accurate representation of the surface brightness profiles of galaxies, which is good enough for the purpose of our study.

The paper is organized as follows. Section~\ref{sec:theprinciple} explains why using {\em the size at a given mass surface density} necessarily reduces the scatter in the size versus $M_\star$ relation.
Section~\ref{sec:math} works out the reduction of scatter to be expected when the surface density profiles are given by Sersic functions. It also points out that using light surface density (i.e., surface brightness) or mass surface density is formally equivalent.
Section~\ref{sec:observations} compares the theoretical predictions with the sizes inferred from the NASA-Sloan Atlas$^{\ref{foot:1}}$ (NSA), which contains parameters for more than 0.6 million galaxies of the local Universe. The theoretical expectations agree extremely well with the observations, and the comparison with one another allows us to infer the mass surface density and surface brightness producing least scatter. Slicing this large catalog also permits studying the dependence of the results on the set of galaxies used in the analysis.
As a by-product of our study, Sect.~\ref{sec:proxy_mass} proposes the use of the inverse size--mass relation to determine $M_\star$.
Finally, Sec.~\ref{sec:discussion} recapitulates and discusses the implications of the work.

%
%
%
\section{The principle}\label{sec:theprinciple}

\begin{figure}
\centering
  \includegraphics[width=0.9\linewidth]{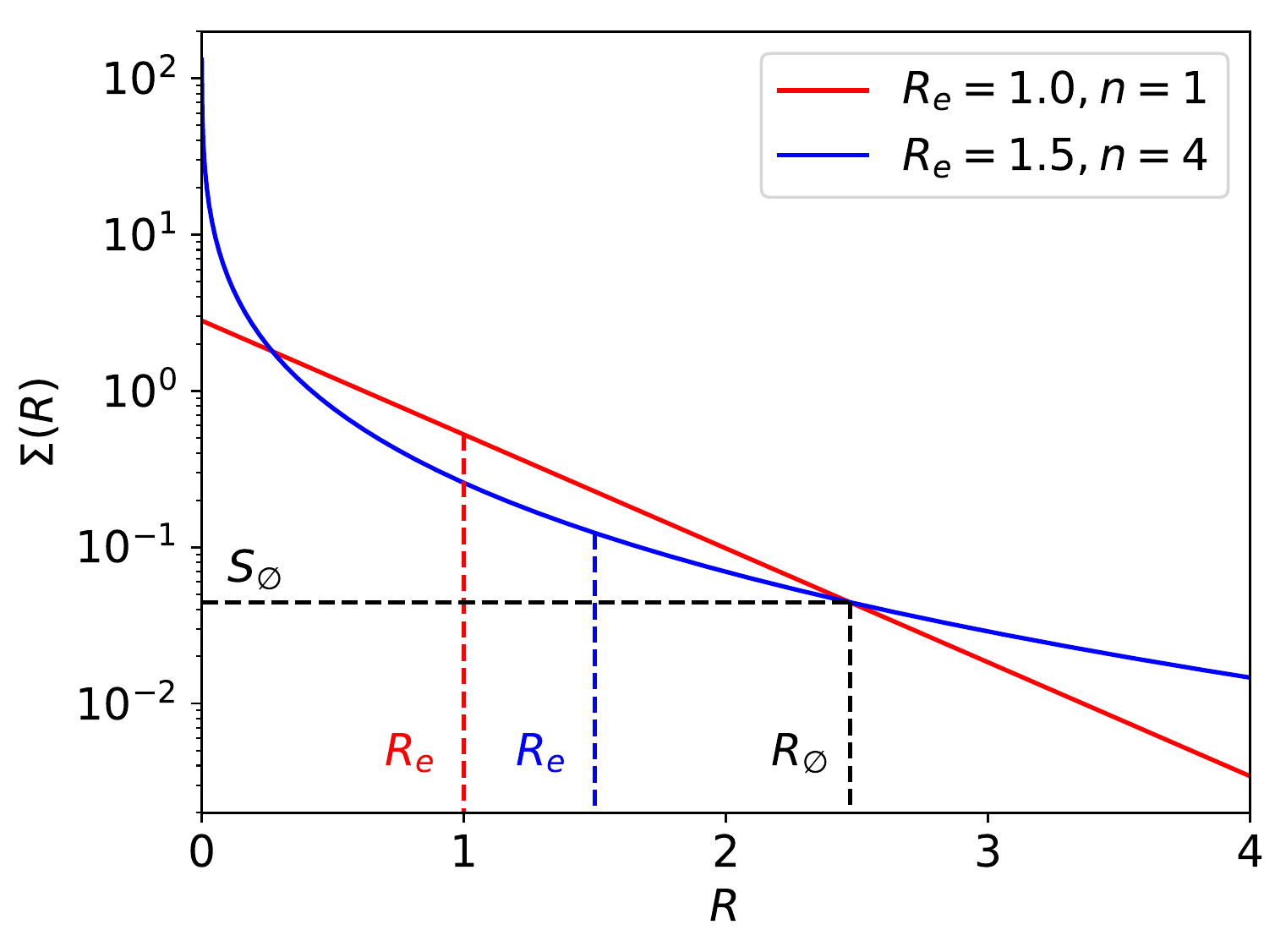}
\caption{Mass surface density profiles for two  galaxies of the same $M_\star$ and different effective radii, $R_e=1$ and $R_e=1.5$. They are assumed to be Sersic functions with the Sersic index given in the inset. The units of $\Sigma$ and $R$ are arbitrary. Because the two profiles have the same normalization, there is always at least one radius $R_\varnothing$ where the two profiles have the same surface density $S_\varnothing$. (In this case there are two such radii.) If the galaxy size is defined as the radius at which the surface density equals $S_\varnothing$ then the two galaxies have exactly the same size.      
}
\label{fig:illustration}
\end{figure}

The stellar surface density of an axisymmetric face-on galaxy can be described in fairly general terms as
\begin{equation}
\Sigma(R) = \frac{M_\star}{2\pi\,w^2}\,f(R/w,\alpha),
\label{eq:sigma}
\end{equation}
where $R$ stands for the distance to the center, $M_\star$ represents the total stellar mass, $w$ is a scale factor, and $\alpha$ corresponds to additional parameters that constrain the density profile. The function $f(x,\alpha)$ gives the shape of the surface density profile, and it is assumed to be normalized,
\begin{equation}
\int_0^\infty\,x\,f(x,\alpha)\,dx = 1.
\end{equation} 
There is a simple mathematical consequence of $f$ being continuous and positive, namely, if two galaxies with two different density profiles have the same stellar mass, then there is at least one radius $R_\varnothing$ where the two density profiles coincide, namely,
\begin{equation}
  \Sigma_1(R_\varnothing) = \Sigma_2(R_\varnothing)= S_\varnothing,
\end{equation}
even when the shape and scale of the profiles of the two galaxies, $\Sigma_1(R)$ and  $\Sigma_2(R)$, are unrelated and very different. If one defines the {\em size} of a galaxy as the radius $R$ where $\Sigma (R) = S_\varnothing$, then the two galaxies have the same size. The idea is illustrated in Fig.~\ref{fig:illustration}.    
Sersic profiles are used in Fig.~\ref{fig:illustration} and throughout the paper since they generally provide a good approximation to the surface density drop in galaxies \citep[e.g.,][]{1968adga.book.....S,1991A&A...249...99C,2005PASA...22..118G}, i.e.,
\begin{equation}
f(R/R_e,n) = \frac{b^{2n}}{n\,\Gamma(2n)}\,\exp\big[-b\,(R/R_e)^{1/n}\big],
\label{eq:sersic}
\end{equation}
where $R_e$ is the mass effective radius (i.e., radius containing half of $M_\star$), $n$ is the so-called Sersic index, and $\Gamma$ is the gamma function. The coefficient $b$ only depends on $n$ and is implicitly given by
\begin{displaymath}
  \Gamma(2n) = 2\,\gamma(2n,b),
\end{displaymath}
with $\gamma$ the incomplete gamma function \citep[e.g.,][]{2005PASA...22..118G}.

\begin{figure*}
\includegraphics[width=0.33\linewidth]{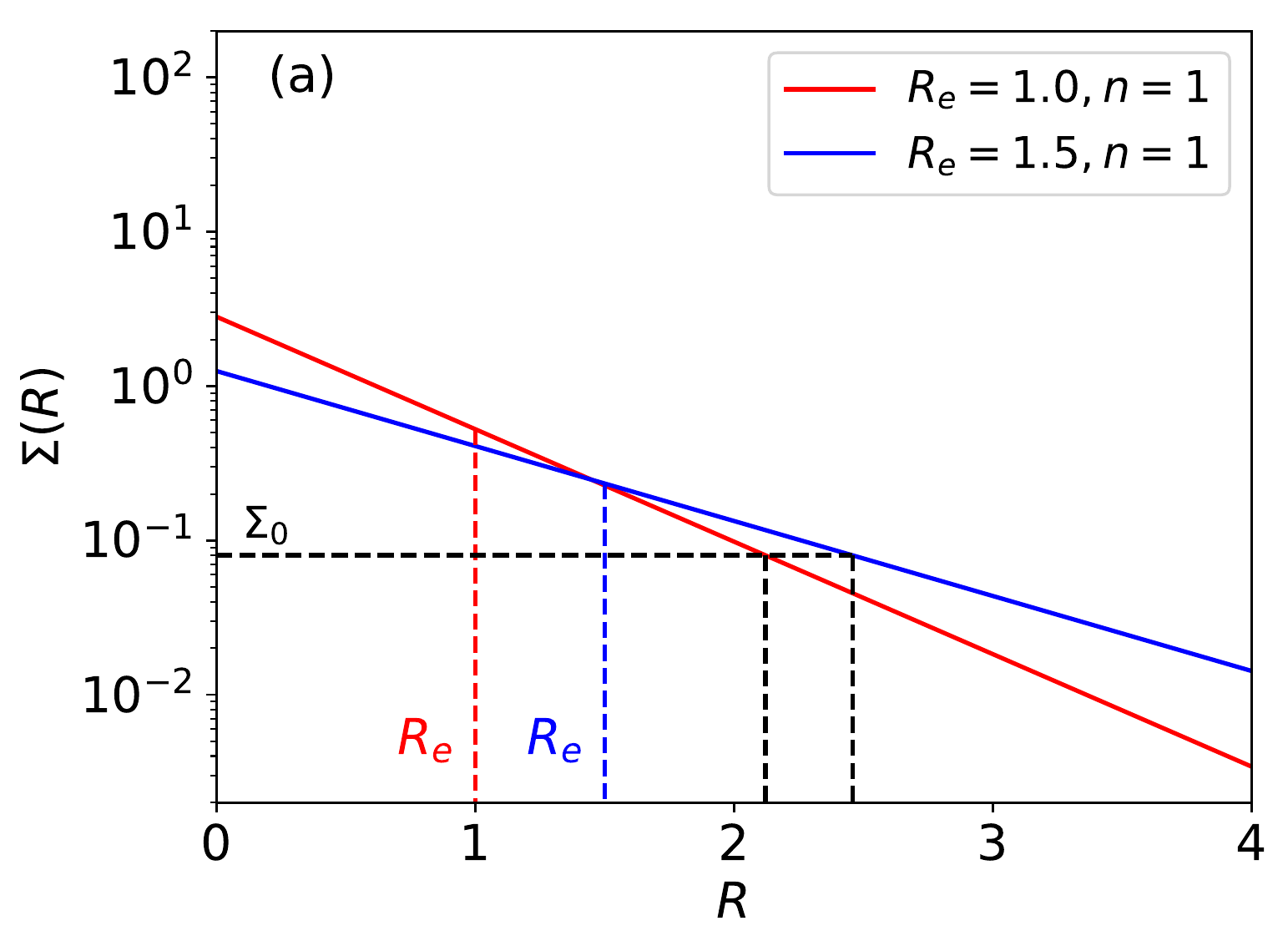}
\includegraphics[width=0.33\linewidth]{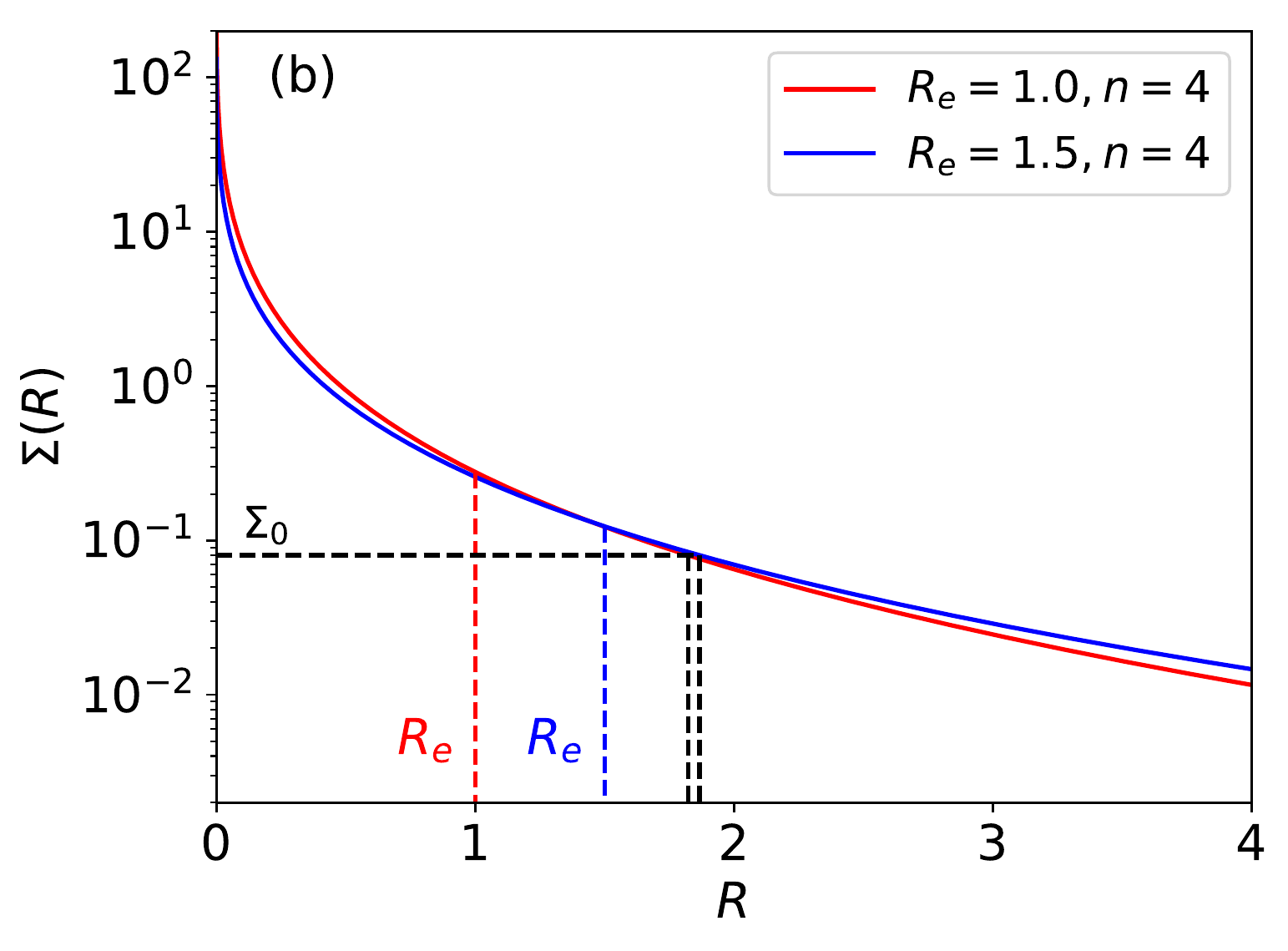}
\includegraphics[width=0.33\linewidth]{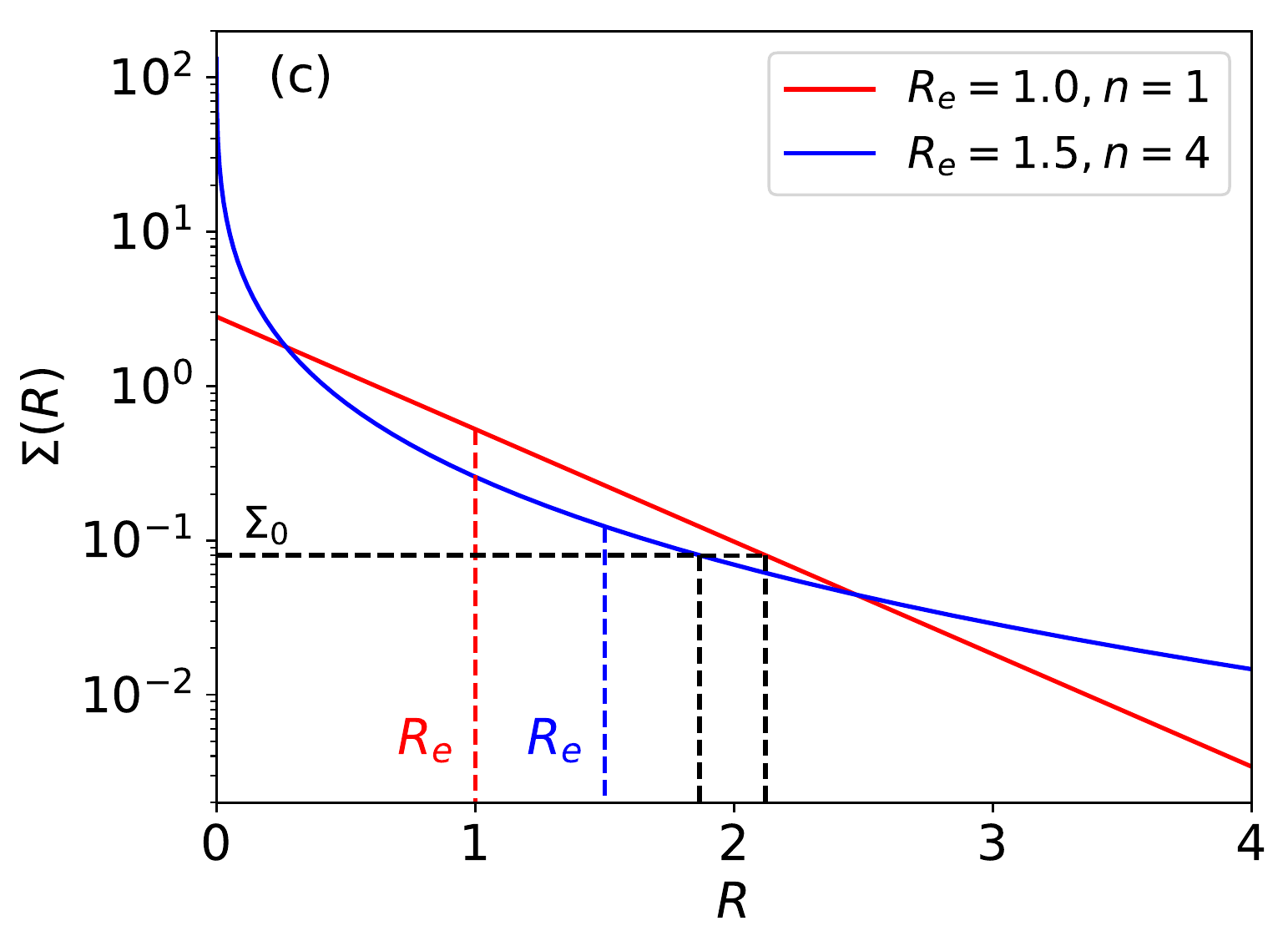}
\caption{(a) Mass surface density profiles for two galaxies of the same $M_\star$ and different effective radii, $R_e=1$ and 1.5. They are assumed to be Sersic functions with the Sersic index given in the inset. The difference between the sizes set by $\Sigma_0$ (shown by vertical dashed black lines) is  smaller than the difference between the effective radii (shown in red and blue as indicated in the inset).  The units of $\Sigma$ and $R$ are arbitrary. (b) and (c) are similar to (a),  including $\Sigma_0$ and $M_\star$. They differ only in the Sersic indexes of the two profiles. 
}
\label{fig:intuition4}
\end{figure*}

Thus, if the size of galaxies  is defined as the radius $R_0$ at which the density profile reaches the value $\Sigma_0$, i.e.,
\begin{equation}
  \Sigma(R_0) = \Sigma_0,
  \label{eq:defsize}
\end{equation}
and if $\Sigma_0$ is chosen close to $S_\varnothing$, then the two above galaxies will not have exactly the same {\em size}, but their difference in size would be smaller than their difference in $R_e$. This is illustrated in Fig.~\ref{fig:intuition4}: the difference between the sizes set by $\Sigma_0$ (shown by vertical black dashed lines) is smaller than the difference between their $R_e$ (shown in red and blue as indicated by the insets in the figure). This happens for a wide range of $n$ and $\Sigma_0$, and thus we argue that this simple principle guides the decrease of scatter in the size-mass relation pointed out by \citet{2020arXiv200102689T}.

\section{Mathematical formulation and general properties}\label{sec:math}

According to the model for $\Sigma$ in Eqs.~(\ref{eq:sigma}) and (\ref{eq:sersic}), the galaxy size defined by Eq.~(\ref{eq:defsize}) is a function of $M_\star$, $R_e$ and $n$, i.e., 
\begin{equation}
R_0 = R_0(M_\star,R_e, n).
\end{equation} 
We will assume that given $M_\star$, the parameters $R_e$ and $n$ are two dependent random variables\footnote{This assumption is quite general and supported by observations, as will be shown in Sect.~\ref{sec:observations}.}. Choosing $R_e$ as the independent variable, 
\begin{equation}
  n = n(R_e).
  \label{eq:n=nre}
\end{equation}
Then applying the law of propagation of errors \citep[e.g.,][]{1971stph.book.....M}, the variance of $R_0$ at a fixed $M_\star$, $\sigma^2_{\log R_0}$, is given by
\begin{equation}
\sigma_{\log R_0} = \Big|\frac{d\ln R_0}{d\ln R_e}\Big|\,\sigma_{\log R_e},
\end{equation}
with $\sigma^2_{\log R_e}$ the variance of the distribution of effective radii.
The larger $|d\ln R_0/d\ln R_{e}|$ the larger the dispersion of sizes compared with the dispersion in $R_e$. Thus, the problem of understanding why the scatter is reduced when using $R_0$ is equivalent to finding when
\begin{equation}
  \Big|\frac{d\ln R_0}{d\ln R_e}\Big|  \leq  1,
  \label{eq:condition}
\end{equation}
and why this happens.

Using the chain rule and  Eqs.~(\ref{eq:sigma}), (\ref{eq:sersic}), (\ref{eq:defsize}), and (\ref{eq:n=nre}), one finds that
\begin{equation}
  \frac{d\ln R_0}{d\ln R_e}=1-\big[\frac{k R_e}{R_0}\big]^{1/n}\,\big[1-\frac{\beta}{2}\,\frac{\partial\ln f}{\partial\ln n}\big],
  \label{eq:fullderiv}
\end{equation}
with
\begin{equation}
  \beta = \frac{d\ln n}{d\ln R_e},
  \label{eq:beta}
\end{equation}
and
\begin{equation}
k =(2n/b)^n\simeq 1.19,
\end{equation}
with the variable $k$ being almost independent of $n$ (it varies from 1.20 to 1.18 for $n$ going from $0.5$ to $6$).

\begin{figure}
  \centering
\includegraphics[width=0.9\linewidth]{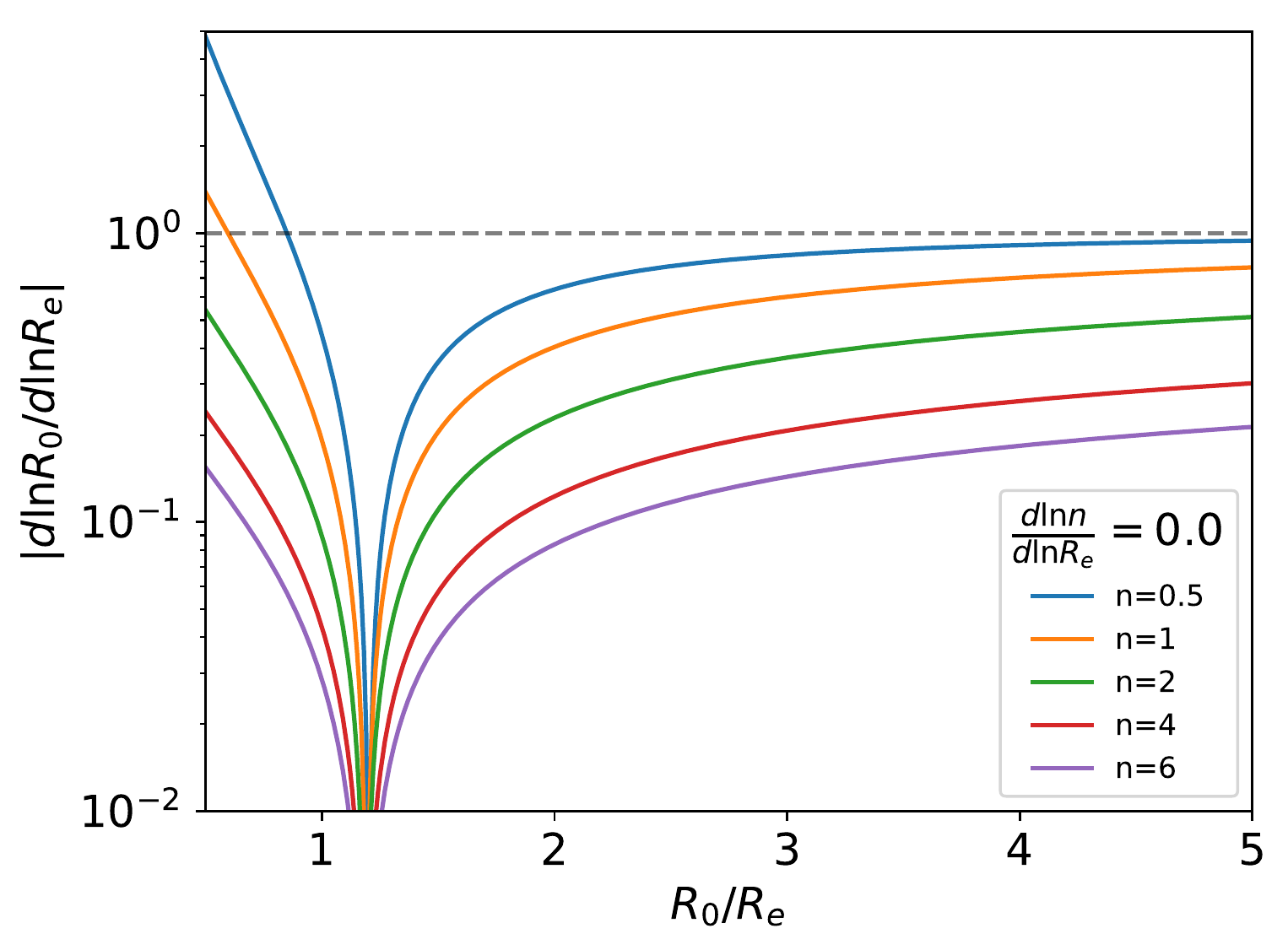}
\caption{
  Dispersion in galaxy size $R_0$  relative to the dispersion in effective radius $R_e$ as a function of $R_0/R_e$. Sizes are defined at a particular surface density (Eq.~[\ref{eq:defsize}]). This case corresponds to $d\ln n/d\ln R_e=0$, so that the Sersic index does not change when $R_e$ varies.  The relative dispersion is almost always smaller than one (set by the horizontal dashed line), and it goes to zero for all Sersic indexes (given in the inset) when $R_0\simeq 1.19\,R_e$. Note also that the  dispersion in $R_0$ is always expected to be smaller for surface density profiles with larger Sersic indexes. $M_\star$ is constant.    
}
\label{fig:intuition5a}
\end{figure}

The case corresponding to constant $n$ (i.e., $\beta = 0$) is particularly simple and illustrative. Equation~(\ref{eq:fullderiv}) yields
\begin{equation}
  \frac{d\ln R_0}{d\ln R_e}= 1-\Big(\frac{k R_e}{R_0}\Big)^{1/n},
  \label{eq:limit}
\end{equation}
and three general results follow seamlessly: (1) for all $n$ and $R_e$, the condition in Eq.~(\ref{eq:condition}) is almost always fulfilled provided $R_0\geq R_e/2$, (2) for all $n$, the dispersion of $R_0$ goes to zero when $R_0\simeq 1.19\,R_e$, and (3) given $R_0/R_e$, the larger the Sersic index the smaller the dispersion in $R_0$.  These three properties are all portrayed in Fig.~\ref{fig:intuition5a}, which shows $|d\ln R_0/d\ln R_e|$ as a function $R_0/R_e$ for $n$ from 0.5 to 6 when $\beta =0$.

\begin{figure*}
  \centering
\includegraphics[width=0.45\linewidth]{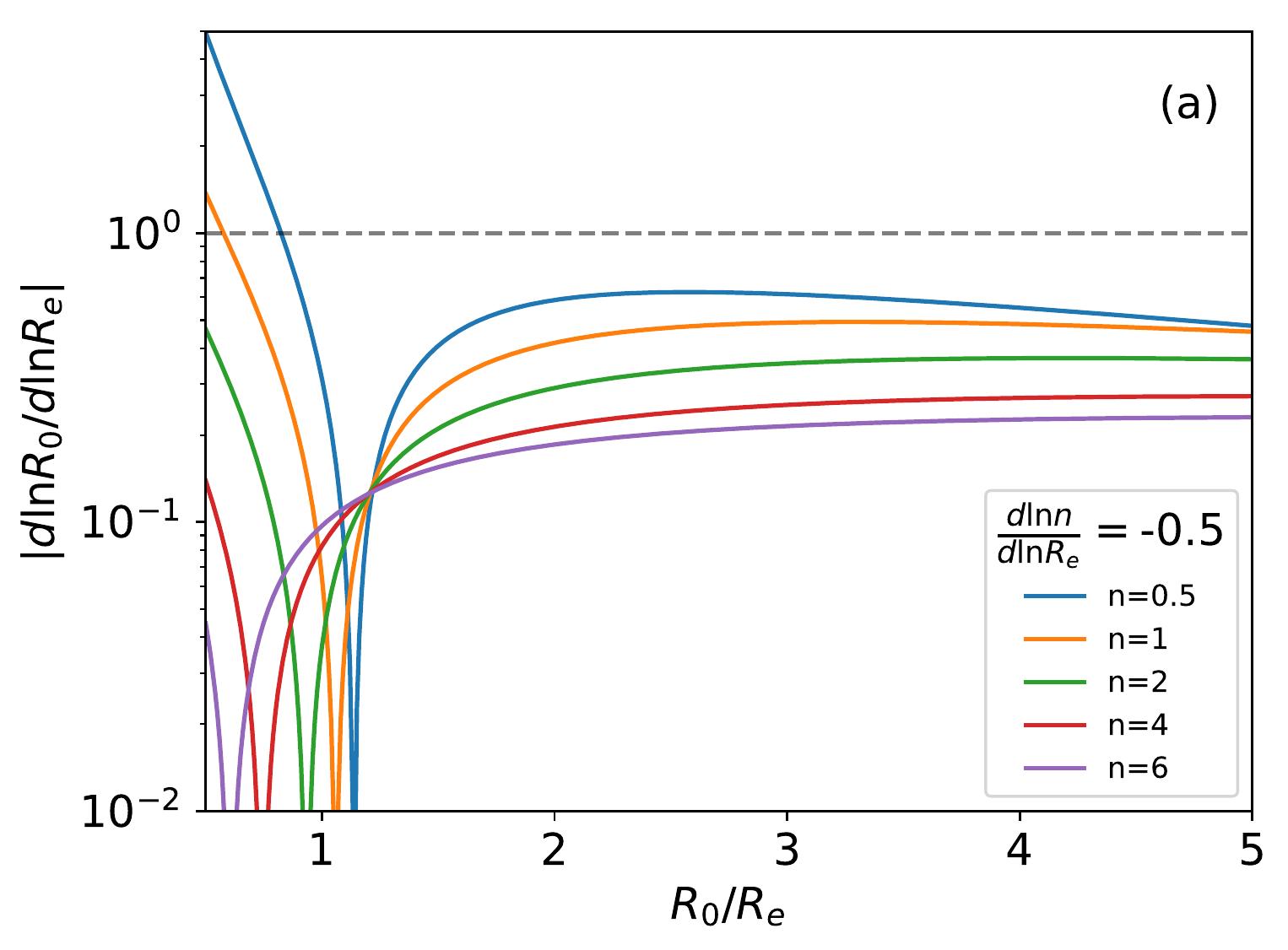}
\includegraphics[width=0.45\linewidth]{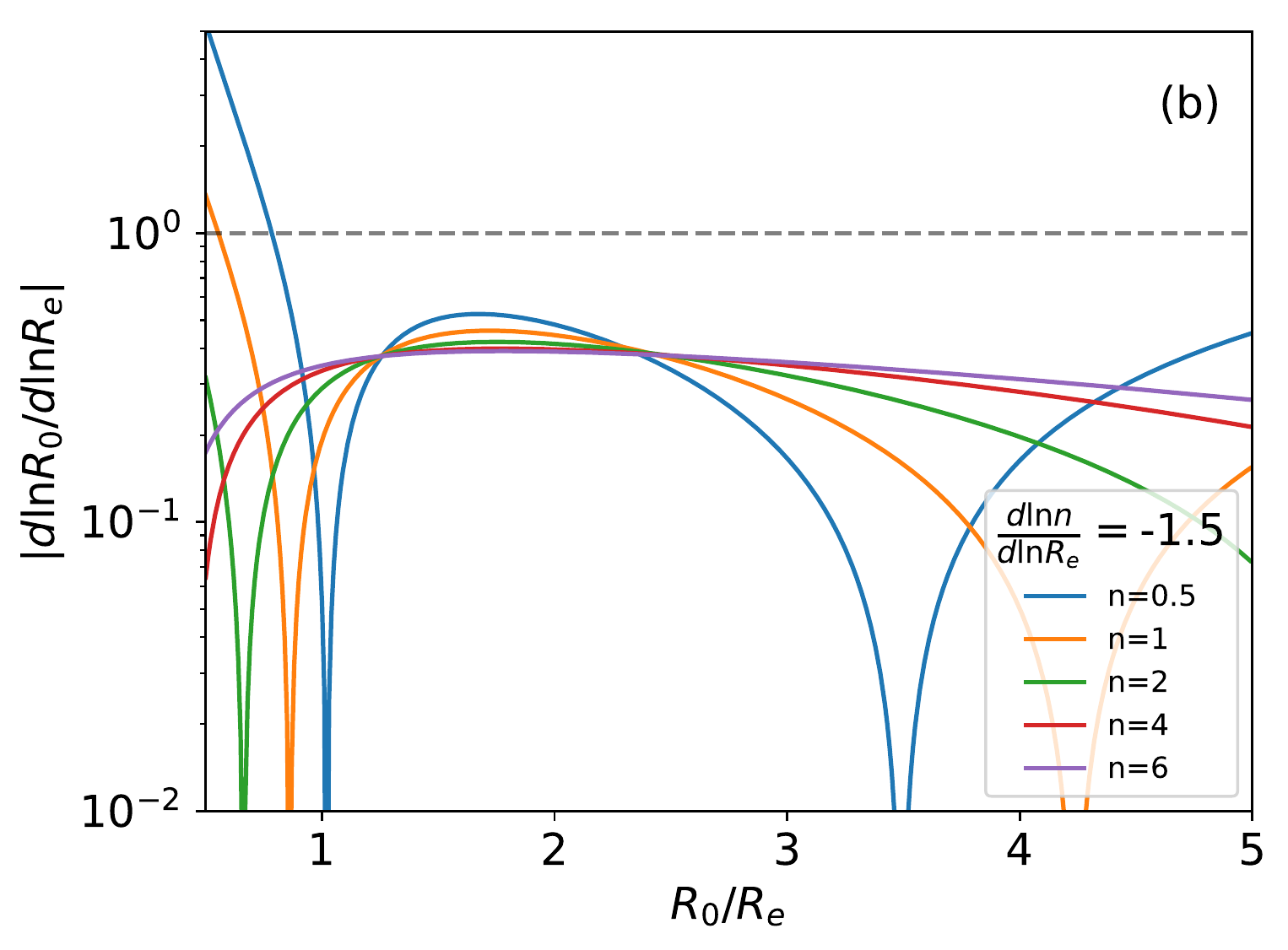}\\
\includegraphics[width=0.45\linewidth]{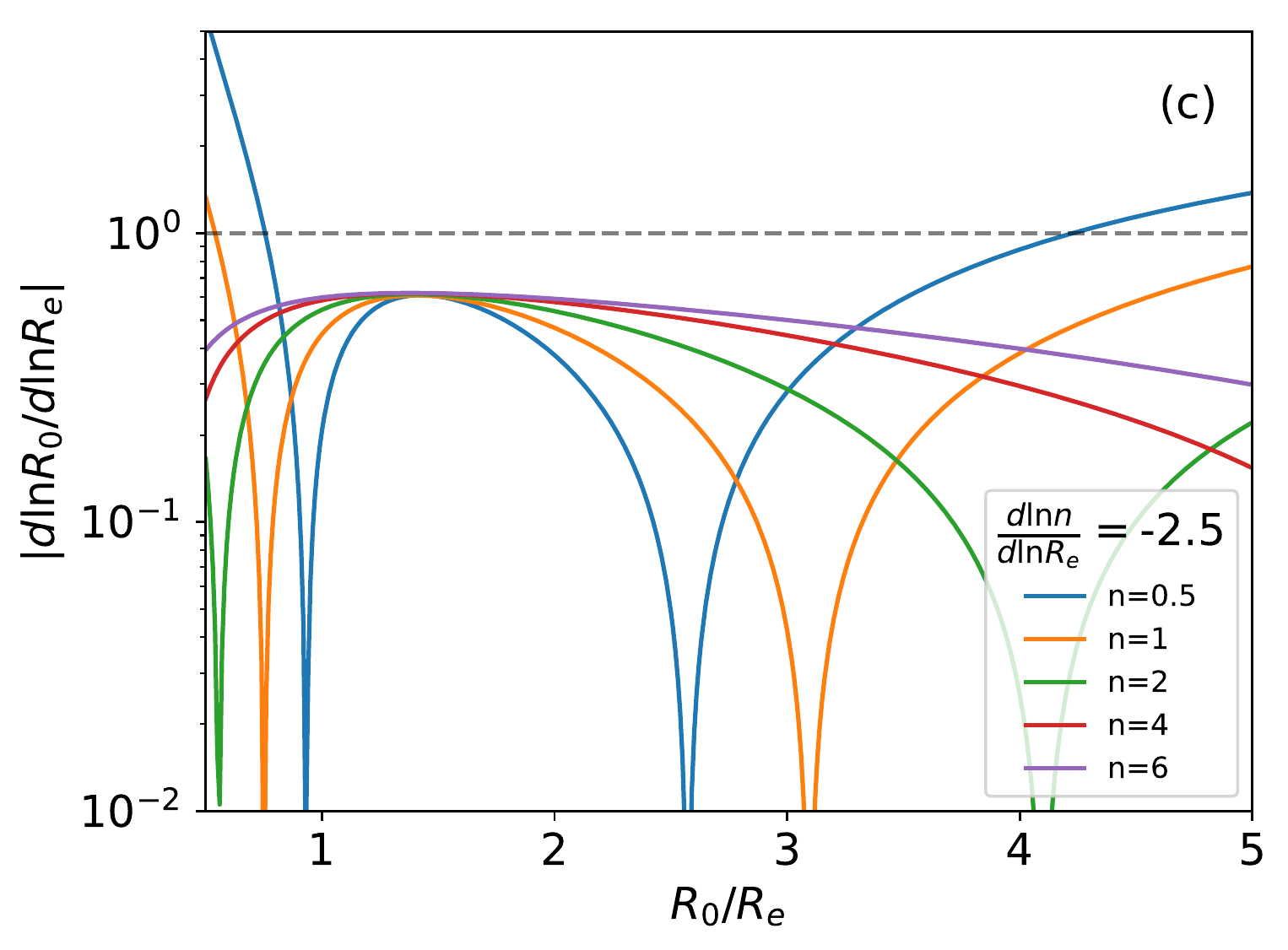}
\includegraphics[width=0.45\linewidth]{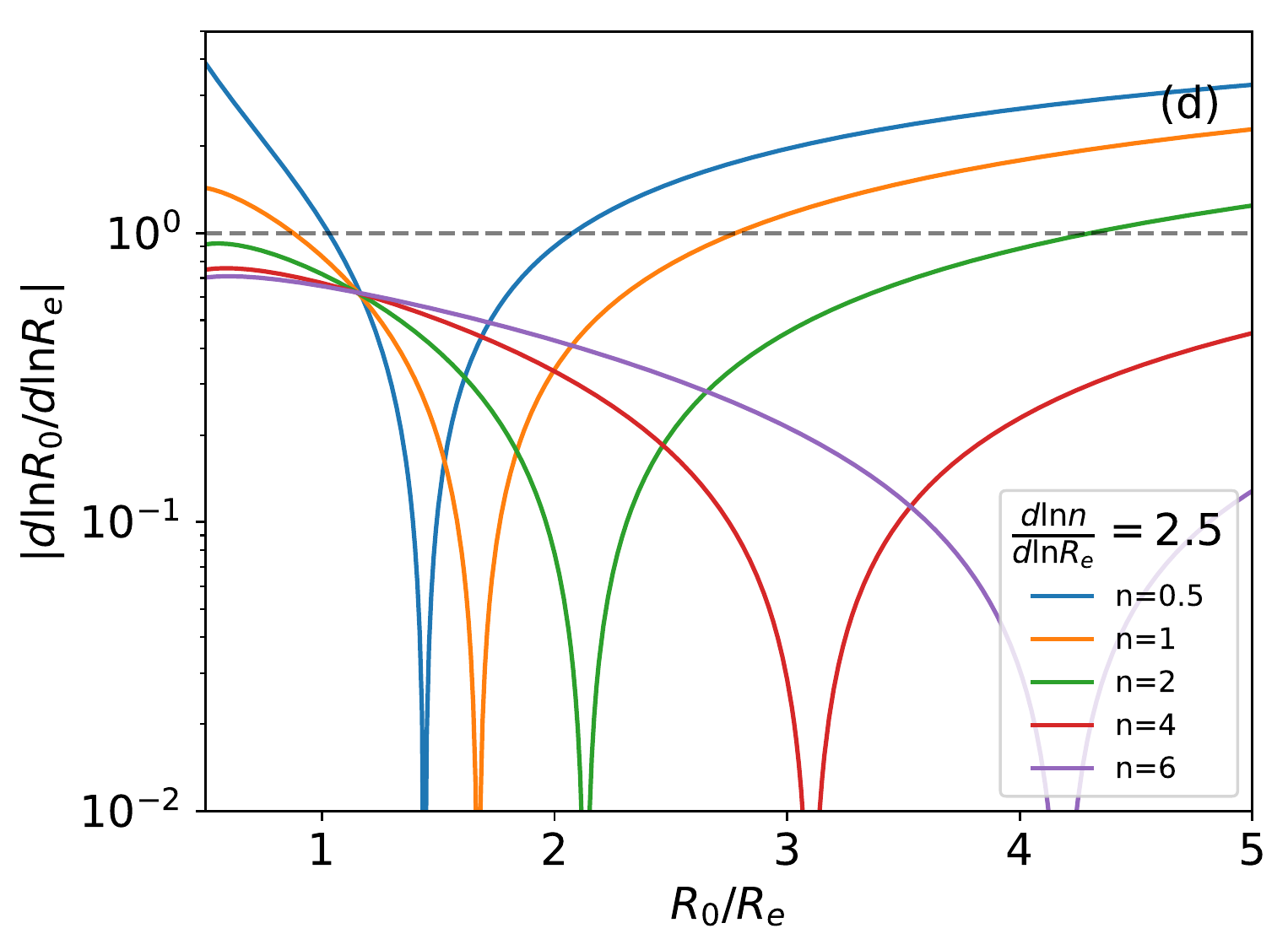}
\caption{
  Dispersion in galaxy size $R_0$  relative to the dispersion in effective radius $R_e$ as a function of $R_0/R_e$. Sizes are defined at a particular surface density (Eq.~[\ref{eq:defsize}]).  The four panels differ only in $d\ln n/d\ln R_e$ as indicated in the insets. It is negative in (a), (b), and (c), and in these three cases, the relative dispersion is almost always smaller than one (set by the horizontal dashed line).  Note also that the spread for different $n$ existing in (a) (and also in Fig.~\ref{fig:intuition5a}) gets largely reduced in (b) and (c) when $1.2 < R_0/R_e < 2.5$. Panel (d) corresponds  to  $d\ln n/d\ln R_e > 0$, and in this case the dispersion is not always  smaller than one. $M_\star$ is  constant in all the plots.
}
\label{fig:intuition5b}
\end{figure*}

The behavior of  $|d\ln R_0/d\ln R_e|$ when $\beta = d\ln n/d\ln R_e$ $\not=0$ is shown in Figs.~\ref{fig:intuition5b}. Panels (a), (b) and (c) correspond to $\beta < 0$, which is the sign observed in galaxies (Sect.~\ref{sec:observations}).  The case of $\beta = -0.5$ (Fig.~\ref{fig:intuition5b}a) is similar to the case of $\beta =0$ analyzed in the previous paragraph (Fig.~\ref{fig:intuition5a}), and will not be discussed any further. However,
a qualitatively different behavior appears when  $\beta < -1$;  for a range of values of $R_0$ around $2\,R_e$,  the dispersion in $R_0$ becomes almost independent of $n$ (Figs.~\ref{fig:intuition5b}b and \ref{fig:intuition5b}c), and yet, the corresponding drop in dispersion between $\sigma_{\log R_0}$ and $\sigma_{\log R_e}$ is quite significant. Finally, Fig. \ref{fig:intuition5b}d portrays a case with $\beta > 0$, showing how the derivatives have grown to the point that $R_0$ often amplifies the scatter existing in $R_e$. As an aside, we note that the derivative $(\partial\ln f)/(\partial\ln n)$ needed to compute  $|d\ln R_0/d\ln R_e|$ in Eq.~(\ref{eq:fullderiv}) has an involved expression in terms of known functions, and so we computed it numerically to produce the curves in Fig.~\ref{fig:intuition5b}.

%
\subsection{Slope of the size -- mass relation}

Applying the chain rule to the model given by Eqs.~(\ref{eq:sigma}),  (\ref{eq:sersic}), and (\ref{eq:defsize}),  one can predict the slope of the size--mass relation when $R_e$ and $n$ are fixed, i.e.,
\begin{equation}
  \frac{\partial\log R_0}{\partial\log M_\star}=\frac{1}{2}\,\Big(\frac{k R_e}{R_0}\Big)^{1/n}.
  \label{eq:slope}
\end{equation}
Since in practice $R_0 > k\,R_e$, the slope of the relation $\log R_0=\log R_0(\log M_\star)$ is expected to increase with increasing  $n$. It also increases when $R_0/R_e$ decreases. Thus, for  $R_0\simeq 2\,R_e$, the slope goes from 0.30 to 0.44 when  $n$ varies from 1 to 4, and it goes from 0.20 to 0.40, when  $R_0\simeq 3\,R_e$. Roughly speaking, this range illustrates the slopes to be expected, even though both $R_e$ and $n$ vary systematically with $M_\star$, which may significantly modify these values.

%
\subsection{Using magnitudes rather than stellar masses}\label{sec:massmageq}
Even if the analysis carried out so far involves mass surface densities, $\Sigma$, it also holds for surface brightness ($SB$). The equations are formally the same replacing $M_\star$ with absolute luminosity, and considering that $SB$ is independent of distance (details are given in App.~\ref{app:appa}). This equivalence is important for the comparison with observations to be carried out in Sect.~\ref{sec:observations}. All observations actually refer to magnitudes and apparent sizes, and it is to be expected that the uncertainties pertaining to transforming them to masses and physical sizes are not present when using $SB$, thus reducing the scatter of the relations even further.

%
\section{Comparison with observations}\label{sec:observations}

\begin{figure}
\includegraphics[width=1.0\linewidth]{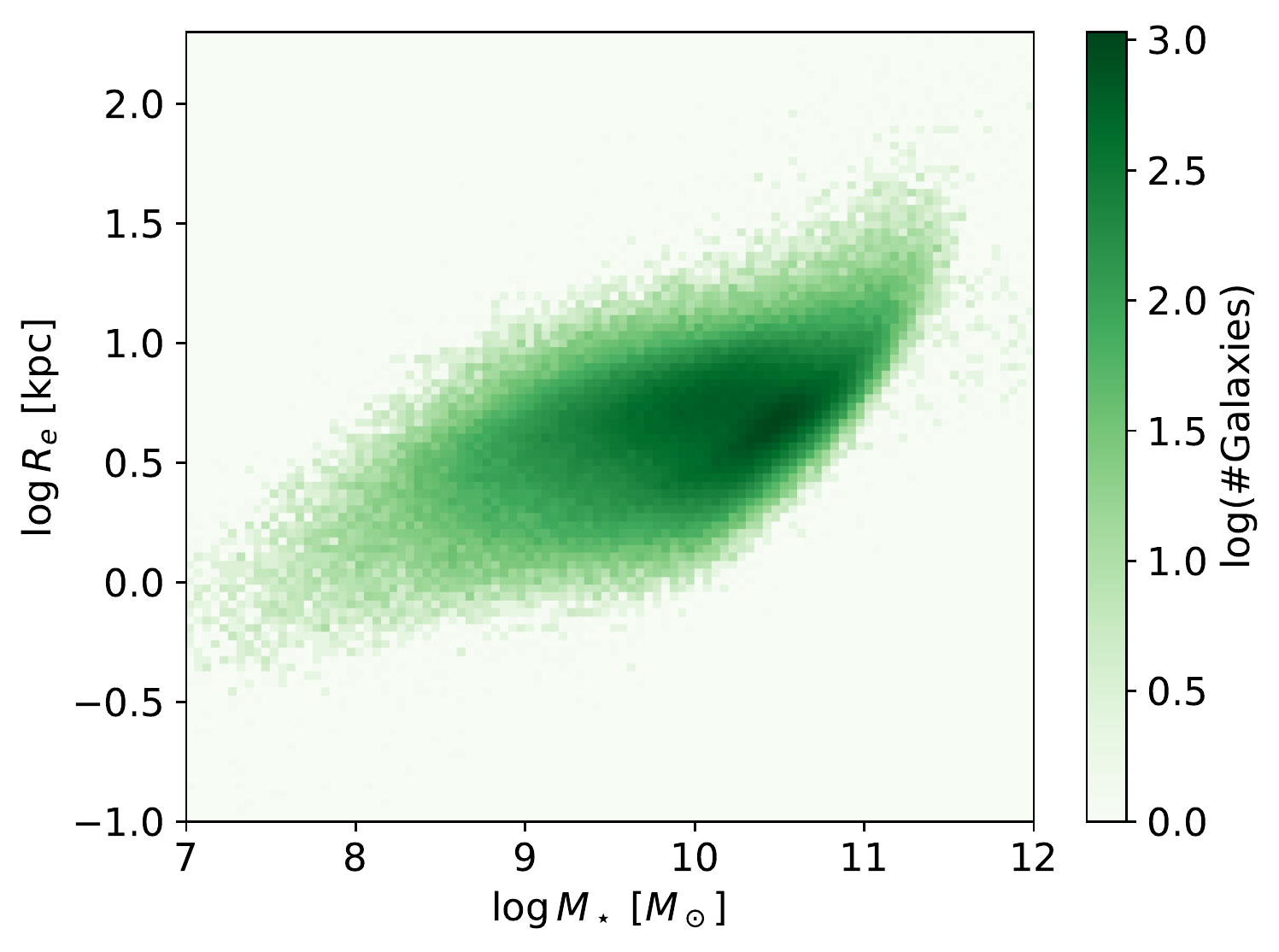}
\includegraphics[width=1.0\linewidth]{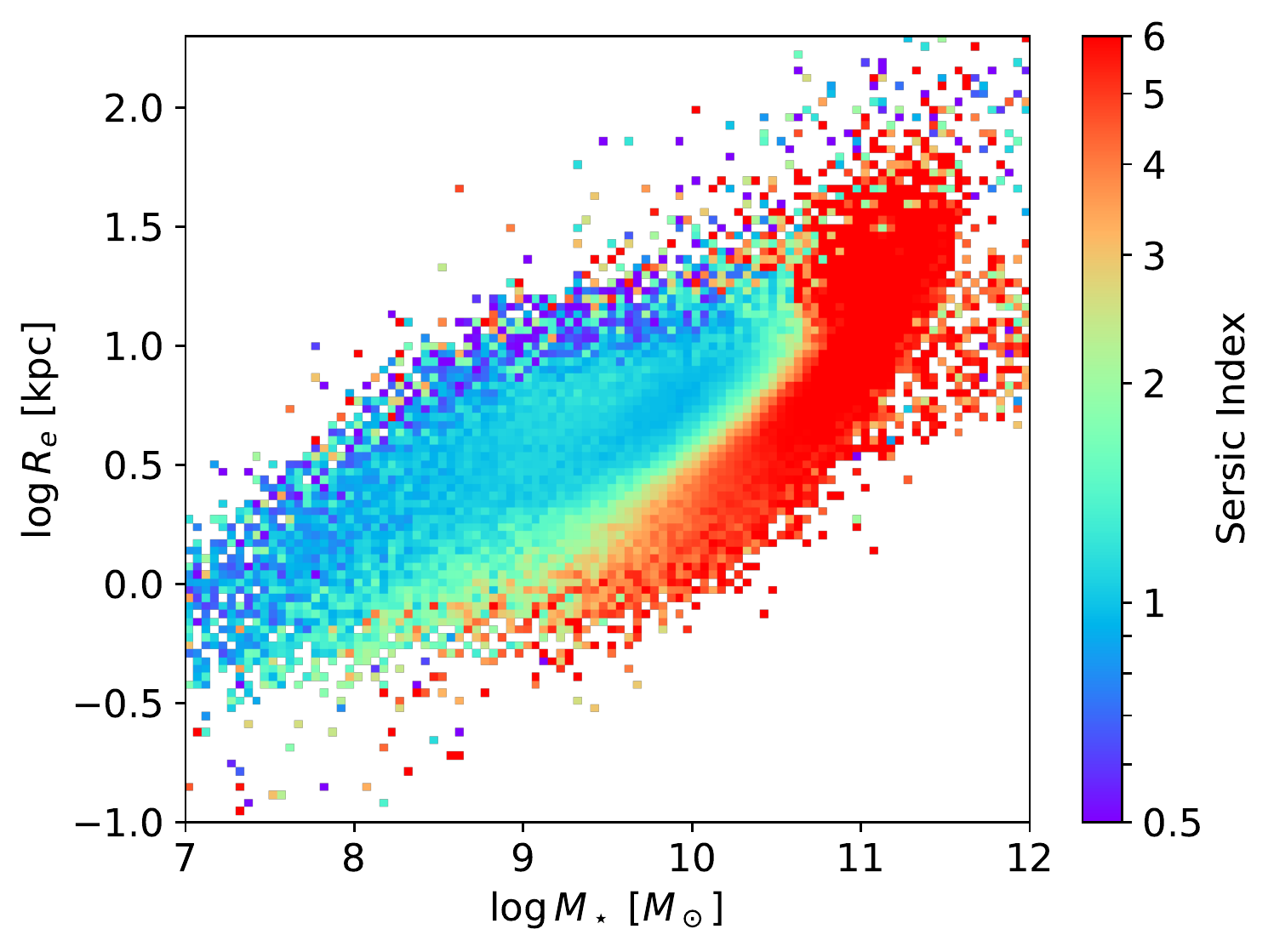}
\caption{
Properties of the galaxies used in our analysis.  Top: Histogram with the number of galaxies in each bin of $\log R_e$ and $\log M_\star$. One glimpses two overlapping distributions that can be identified with the blue cloud  (having larger $R_e$ for their mass) and the red sequence. 
Bottom:  Same plane as above color-coded according to the median of the Sersic index of the galaxies in the bin. For a fixed  $M_\star$, galaxies in the blue cloud have larger $R_e$ and smaller $n$ than the galaxies of the red sequence.
}
\label{fig:getsizes4}
\end{figure}

\begin{figure}
  \centering
\includegraphics[width=.9\linewidth]{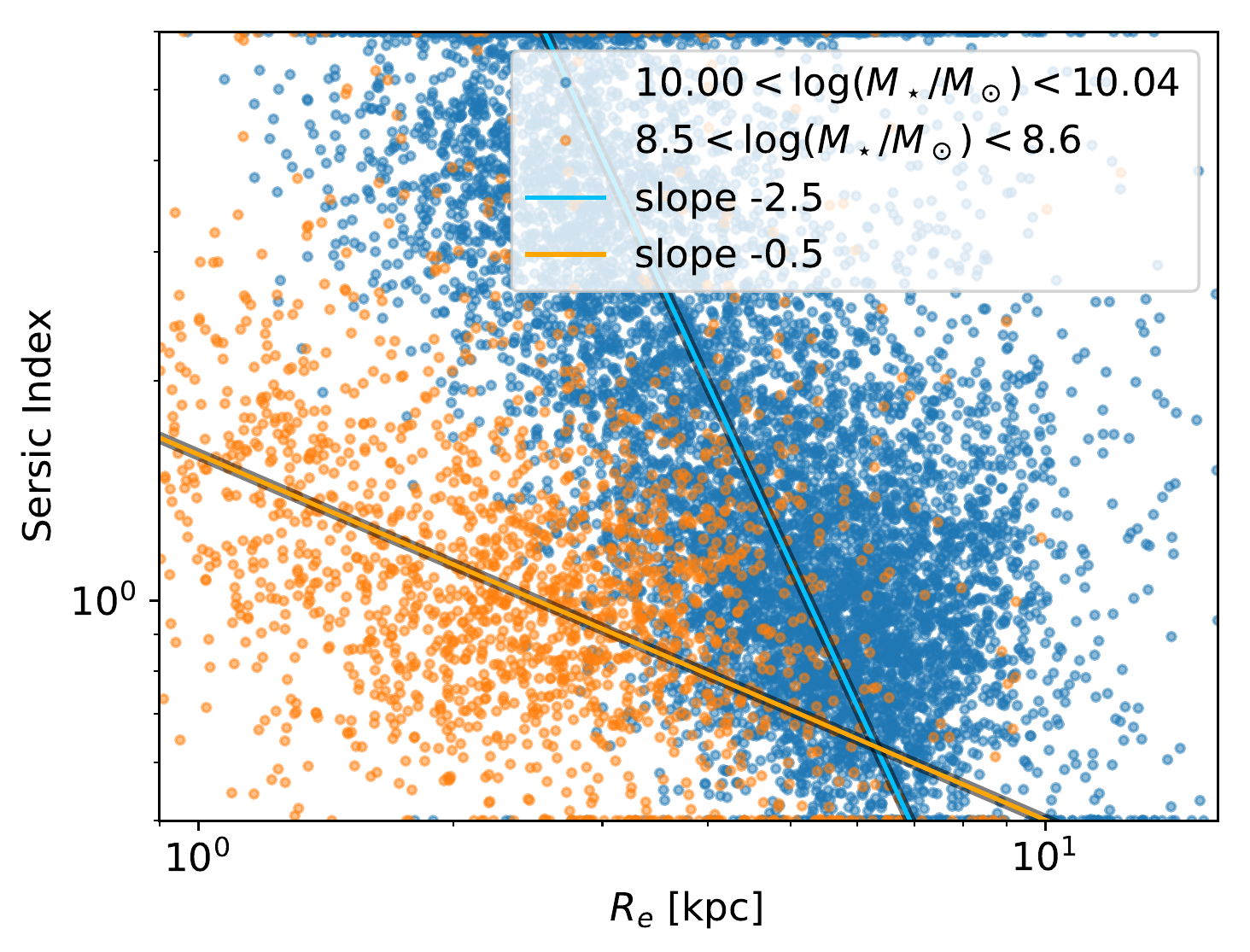}
\caption{
    Sersic index $n$ versus effective radius $R_e$ for galaxies in the NSA catalog with $M_\star$ within the two narrow ranges indicated in the inset. $n$ and $R_e$ are anti-correlated with a slope  in the log--log representation going from $-0.5$, at lower $M_\star$, to $-2.5$, at higher $M_\star$.  In the parlance of Sect.~\ref{sec:math}, $n\propto R_e^\beta$ (Eqs.~[\ref{eq:n=nre}] and [\ref{eq:beta}])  with $\beta$ going from  $-2.5$ to $-0.5$ depending on $M_\star$.}
\label{fig:getsizes2_1}
\end{figure}

The properties of the model surface density profiles worded out in Sect.~\ref{sec:math} match very well the behavior observed in galaxies. The NASA-Sloan Atlas\footnote{\label{foot:1}{\tt  http://nsatlas.org}}  (NSA) represents a convenient testbed to support this claim since the catalog  provides all the physical parameters required for the tests, and it contains a very large number of local galaxies. The NSA catalog provides $R_e$, $n$, absolute magnitudes $M_r$, distances, and $M_\star$ of almost all SDSS-DR11 galaxies \citep[][]{2015ApJS..219...12A} with know redshifts out to $\sim 0.15$. $R_e$ and $n$ are inferred from the $SB$ profile as described by \citet{2011AJ....142...31B}, and so they refer to the photometry of the galaxy.
The determination of $R_e$ and $n$ employs elliptical apertures, with the position angle and axis ratio of the ellipses inferred from the second moments of the light distribution. This approach corrects for the inclination of the galaxies except in pathological cases \citep[e.g.,][]{2019RNAAS...3..191S}, providing $n$ and $R_e$ directly comparable to the face-on galaxies modeled in Sect.~\ref{sec:math}. 
$M_\star$ was inferred assigning a mass-to-light ratio to the observed broad-band colors. The estimate assumes as initial-mass-function (IMF) the {\em diet} Salpeter IMF described by \citet{2003ApJS..149..289B}, which yields $M_\star$ slightly larger than other common IMFs \citep[$\sim 0.15\,$dex; ][]{2003ApJS..149..289B,2009ARA&A..47..159B}.
The NSA catalog contains about 640,000 galaxies, and it was trimmed by removing galaxies with significant error in magnitude and effective radii. We asked for relative error in magnitude $< 1$\,\%, similar effective radius in $g$ and $r$ (5\,\%), and large enough apparent effective radius ($>$ 1.5 arcsec). (None of these cuts modify the results discussed below -- they merely reduce the noise in the empirical relations.) The cleaning reduces a factor of two the sample size, leaving around 290,000 galaxies. Since $n$ in NSA is computed in the $r$-band, we use magnitudes and effective radii in the $r$-band as well. The galaxies thus selected are used to produce Fig.~\ref{fig:getsizes4}. The top panel displays a 2D histogram with the number of galaxies in each bin of $\log R_e$ and $\log M_\star$. Two clumps are identifiable, which can be ascribed to the blue cloud and red sequence, or to late-type and early-type galaxies \citep[e.g.,][]{2009ARA&A..47..159B}.  The bottom panel in Fig.~\ref{fig:getsizes4} shows the location of the galaxies in the same plane $\log R_e$ versus $\log M_\star$, color-coded according to the median Sersic index in the bin. Overall, more massive galaxies have larger both $R_e$ and $n$. However, for a fixed stellar mass, galaxies in the blue cloud have larger $R_e$ and smaller $n$ than the galaxies in the red sequence, a characteristics already reported in the literature \citep[e.g.,][]{2003MNRAS.343..978S,2015MNRAS.447.2603L}. 
Figure~\ref{fig:getsizes4}, bottom panel, clearly shows an increase of $n$ with decreasing $R_e$ at a given $M_\star$.  It is more evident in  Fig. \ref{fig:getsizes2_1}, which contains  the scatter plot $n$ versus $R_e$ in two narrow ranges of $M_\star$. The observed points show a significant anti-correlation which, in  the parlance of Sect.~\ref{sec:math},  implies that $\beta$  (Eq.~[\ref{eq:beta}]) goes from $-2.5$ to $-0.5$ depending on $M_\star$.

As we discuss in Sect.~\ref{sec:massmageq}, even if the analysis in Sect.~\ref{sec:math} is carried out for mass density profiles to comply with the work of \citet{2020arXiv200102689T}, it also holds for surface brightness ($SB$). The equations are formally the same (App.~\ref{app:appa}), which  allows us to use the NSA catalog with $M_r$ and $SB$ to test the theoretical results worked out in Sect.~\ref{sec:math}. The comparison will be then extended to $\Sigma$ and $M_\star$ in Sect.~\ref{sec:alsomass}.

\begin{figure*}
  \centering 
\includegraphics[width=0.9\linewidth]{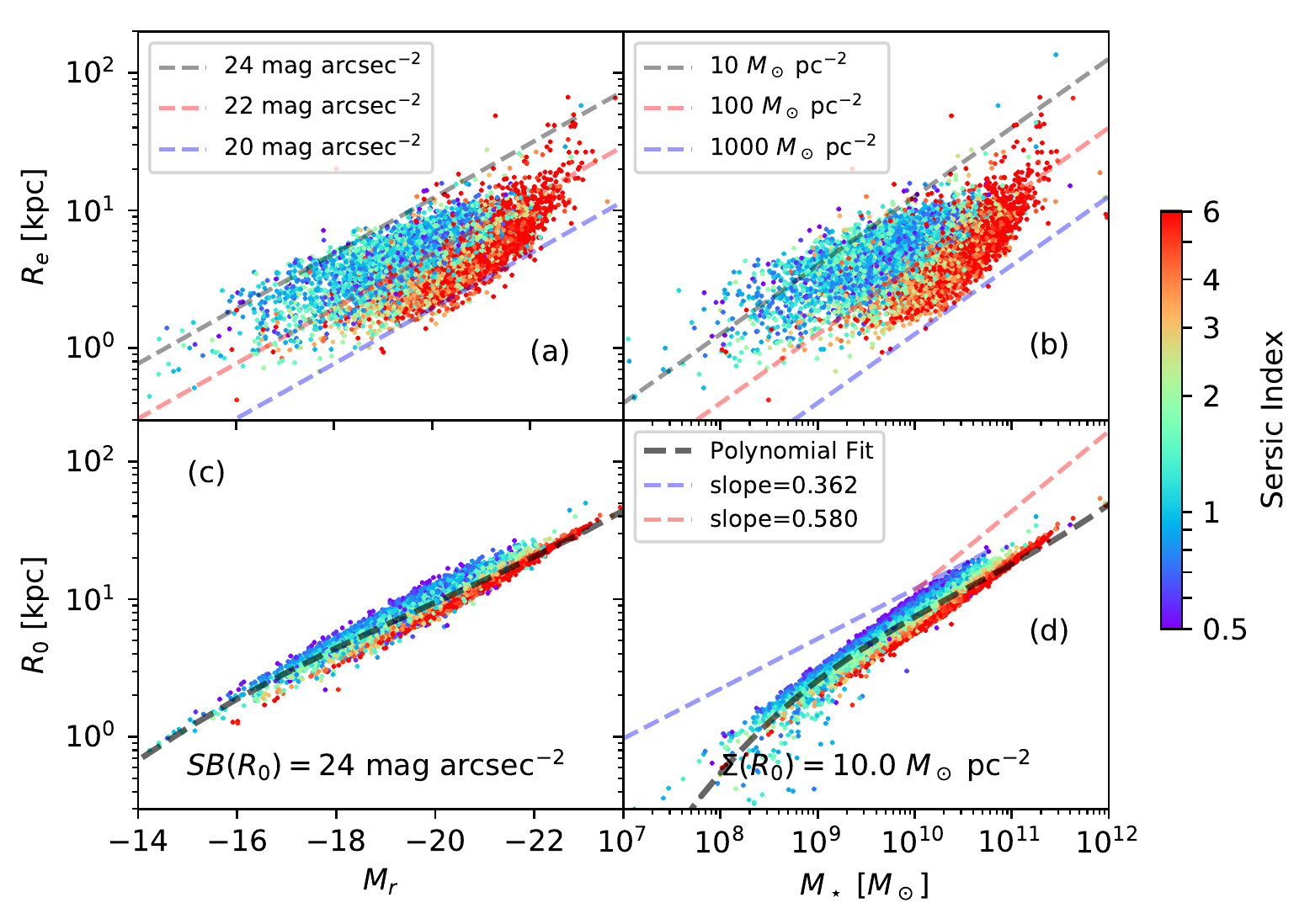}
\caption{
Top panels: (a) Effective radius $R_e$ versus absolute magnitude $M_r$ and (b) $R_e$ versus stellar mass $M_\star$ for some 8800 galaxies randomly chosen from the NSA catalog ($\sim$3\,\% of the sample). The lines represent the sequence followed by objects with the same average half-light surface brightness (a) and half-mass surface brightness (b), with the constant values indicated in the insets. 
Bottom panels: Same as upper panels with the size of the galaxy $R_0$ inferred at fixed surface density, either in luminosity (c) or in mass (d). The used $SB_0$ and $\Sigma_0$ are given in the corresponding panel.  The black dashed lines are 3rd order polynomial fits to the scatter plots. The blue and red lines in panel (d) are used for reference and correspond to the sequence followed by the galaxies analyzed in \citet{2020arXiv200102689T}.   Points in all four panels are color-coded according to $n$, as indicated by the color bar.
}
\label{fig:getsizes0}
\end{figure*}

Figure~\ref{fig:getsizes0}a shows the scatter plot $R_e$ versus $M_r$ for the NSA galaxies color-coded with $n$. In order to avoid overcrowding, only 3\,\% of the galaxies in the sample were chosen for  representation ($\sim 8800$ galaxies). (Other random selections give very similar results.)  Figure~\ref{fig:getsizes0}c shows $R_0$ versus $M_r$ for an arbitrarily chosen constant $SB_0$ (in this case 24  mag~arcsec$^{-2}$) so that
\begin{equation}
SB(R_0) = SB_0.
\end{equation}
As detailed  in App.~\ref{app:appa}, $R_0$ was inferred from the observables assuming the light profile to follow a Sersic function. The relation between $R_0$ and $M_r$ clearly shows all the properties mentioned in Sect.~\ref{sec:math}: (1) the scatter  is reduced with respect to the use of $R_e$ (cf. Figs.~\ref{fig:getsizes0}a and \ref{fig:getsizes0}c). (2) This reduction depends very much on the Sersic index. In this particular case, the scatter is smaller for larger Sersic index, at it is expected from  Fig.~\ref{fig:intuition5b}c because $R_0\sim$\,2--3\,$R_e$ and $\beta \sim -2.5$ (Fig.~\ref{fig:getsizes2_1}). (3) The slope of the relation between $\log R_0$ and $M_r$  is also in the range given by Eq.~(\ref{eq:slope}) (keeping in mind the factor 2.5 involved in the definition of magnitudes).  (4) Figure~\ref{fig:getsizes0}b uses $SB_0=24\,$mag\,arcsec$^{-2}$, but the qualitative behavior it leads to is representative of other values as well, as we elaborate on in Sect.~\ref{sec:least}. 

\subsection{Sizes inferred from mass surface densities}\label{sec:alsomass}
If the mass-to-light ratio of each galaxy is assumed to be constant then the values for $R_e$ and $n$ inferred from photometry are also valid for the mass surface density profile. Under this assumption, we also derive $R_0$ from the observed $M_\star$, $R_e$ and $n$ (Eqs~[\ref{eq:sigma}], [\ref{eq:sersic}], and [\ref{eq:defsize}]).  The observed scatter in the relation $R_e$ versus $M_\star$ (Fig.~\ref{fig:getsizes0}b) narrows down in Fig.~\ref{fig:getsizes0}d when $R_0$ is inferred via Eq.~(\ref{eq:defsize}). Figure~\ref{fig:getsizes0}d employs $\Sigma_0=10\,M_\odot\,$pc$^{-2}$, but most properties are not specific of the actual value. As it happened with the sizes inferred from $SB$, the scatter depends on $n$ and is smaller for larger values.  We will show in Sect.~\ref{sec:least} that the scatter is reduced even further using a particular value for $\Sigma_0$, and then the resulting relation becomes very close to the relation found by \citet{2020arXiv200102689T}, which do not rely  on the approximations used here.  Namely, their $R_e$ and $R_0$ are directly derived from the observed mass surface density without assuming Sersic profiles. The agreement between their relation and ours allows us to argue that the assumption on $\Sigma$ following a Sersic shape does not hamper our analysis (see also Sect.~\ref{sec:discussion}).

\subsection{Surface brightness producing least scatter in $R_0$}\label{sec:least}
\begin{figure}
  \centering
\includegraphics[width=0.7\linewidth]{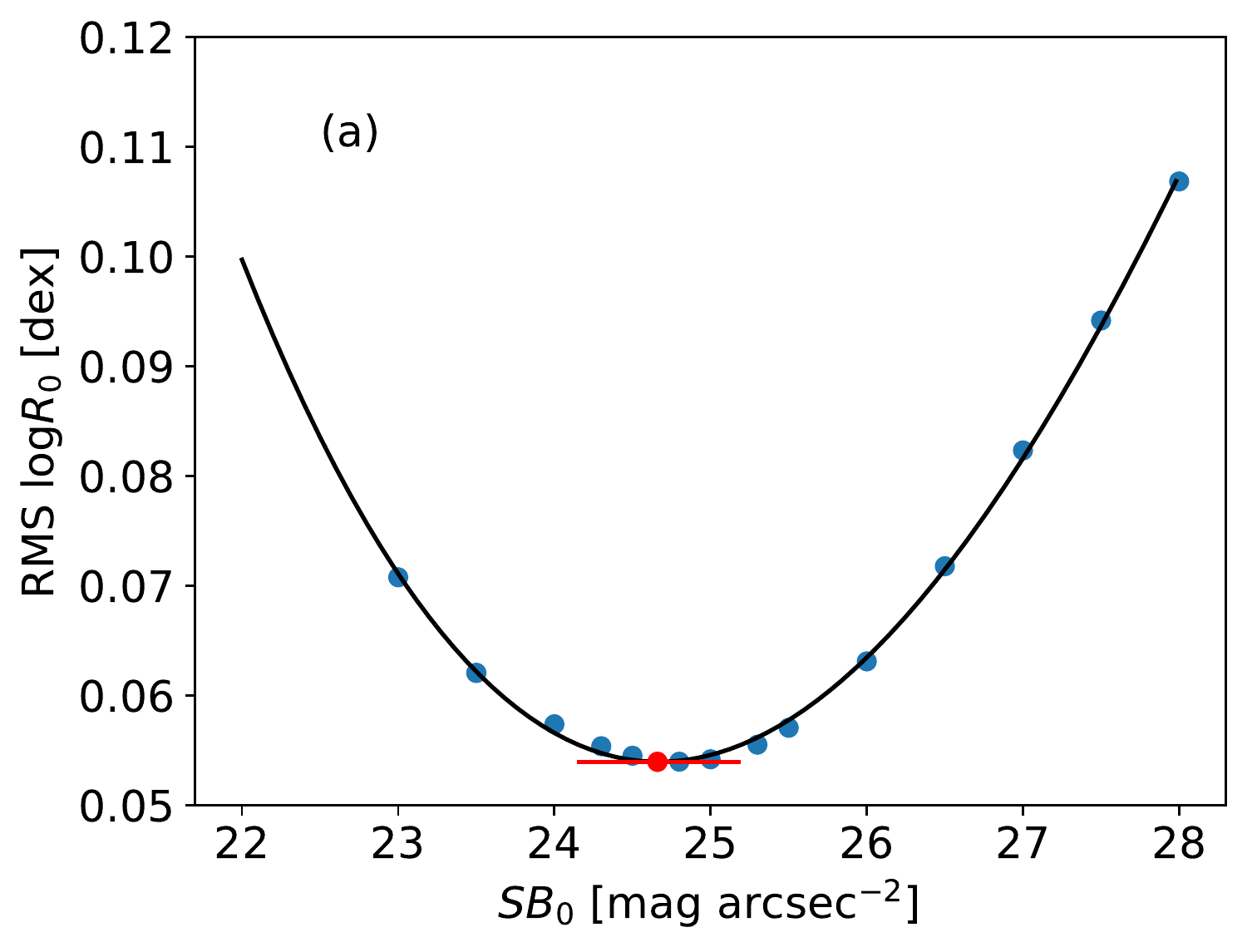}
\includegraphics[width=0.7\linewidth]{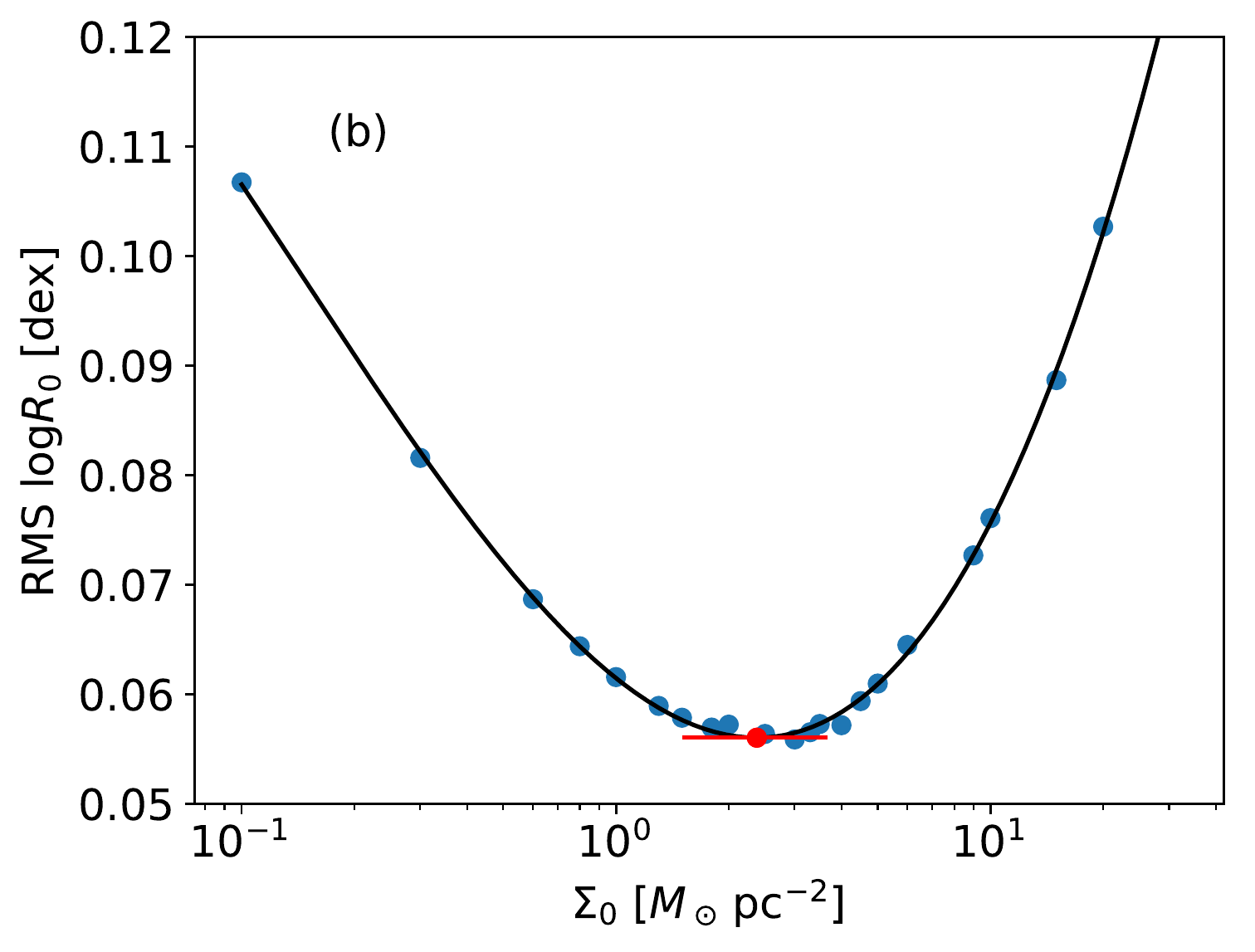}
\caption{
(a)  RMS of the relation $\log R_0=\log R_0(M_r)$ as a function of $SB_0$, i.e., the surface brightness defining the size. The least scatter in the relation is achieved for $SB_0\simeq 24.7\pm 0.5$ mag\,arcsec$^{-2}$ (the range is given by the red horizontal bar, and it refers to AB magnitudes in the $r$-band). The blue symbols represent actual observations whereas the solid line shows the 3rd order polynomial fit used to locate the minimum RMS (the red symbol). (b) Same as (a) but referring to the search for the optimal $\Sigma_0$, which turns out to be $2.4_{-0.9}^{+1.3}\,M_\odot\,{\rm pc}^{-2}$ as indicated by the red symbol and horizontal line. 
}
\label{fig:optimum}
\end{figure}

As it is clear from the theoretical analysis carried out in Sect.~\ref{sec:math}, the scatter of the relation $R_0=R_0(SB_0)$  (and so of $R_0 =R_0[M_\star]$) depends on $SB_0$ (on $\Sigma_0$). The question arises as to what is the value of $SB_0$ producing least scatter. We approach the problem fitting a smooth function to the relation $\log R_0=\log R_0(M_r)$ (a 3rd order polynomial suffices) and then computing the root mean square (RMS) of the residuals. This is repeated for a range of $SB_0$ values, and the resulting RMS versus $SB_0$ is shown in Fig.~\ref{fig:optimum}a.  We find the minimum to be at $SB_0\simeq 24.7\pm 0.5$ ($r$-band; AB system). At this minimum, the RMS of the relation is just 0.054\,dex. The error bar gives the range of $SB_0$ that provide RMS values compatible with the residuals of the fit (see Fig.~\ref{fig:optimum}).

The same exercise repeated  with $\log R_0=\log R(\log M_\star)$ leads to an optimal $\Sigma_0=2.4_{-0.9}^{+1.3}\,M_\odot$\,pc$^{-2}$, for which the minimum RMS is 0.056\,dex, and thus slightly higher than the value inferred from $SB$. The difference may be in the errors to compute $M_\star$ and on the implicit assumption needed to use for $\Sigma$ the parameters $n$ and $R_e$ derived from photometry. As we argue below, the two optimal values are very much consistent with each other.

\begin{figure*}
  \centering
\includegraphics[width=0.9\linewidth]{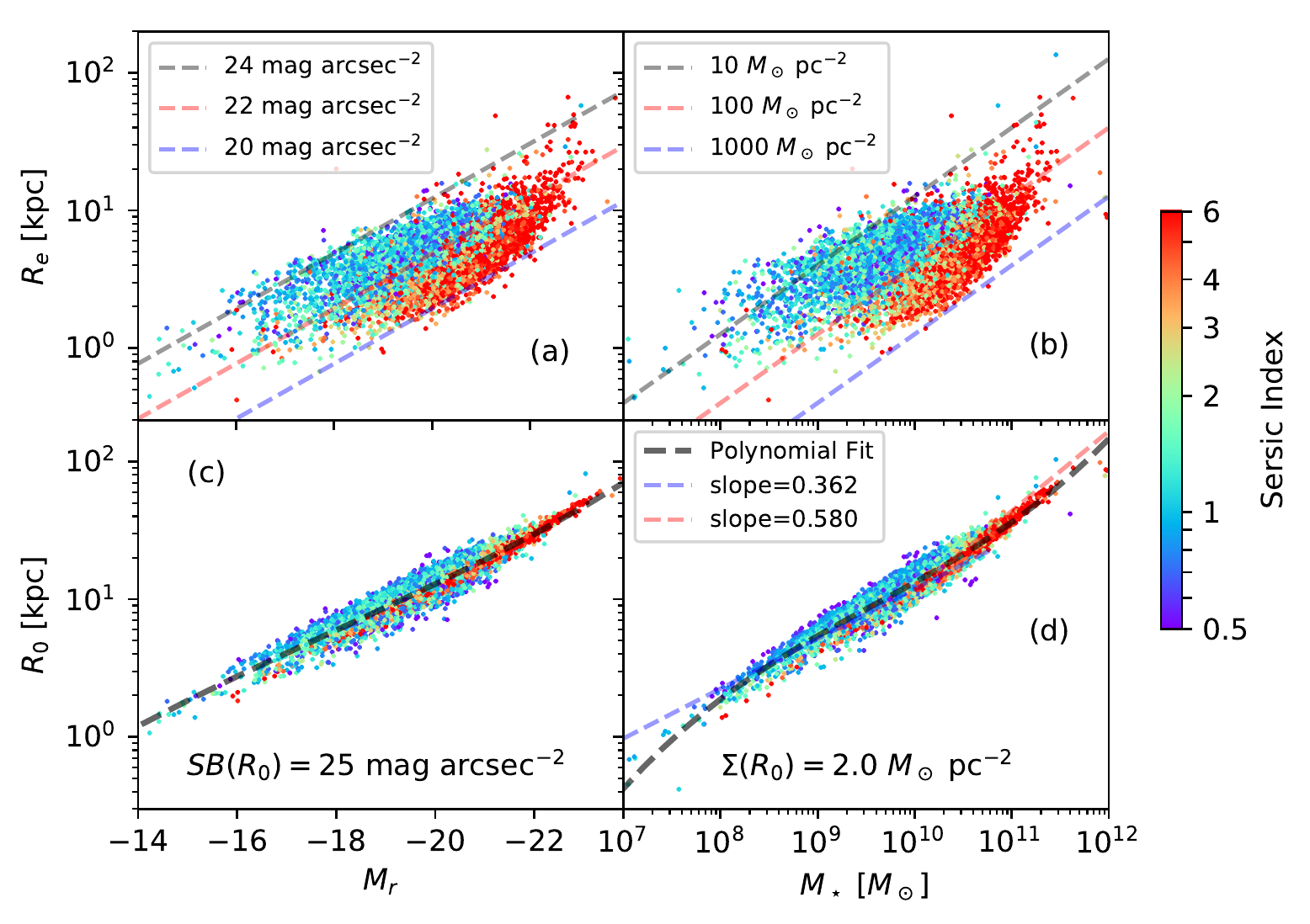}
\caption{
Same as Fig.~\ref{fig:getsizes0}, except for the values of $SB_0$ and $\Sigma_0$.  The values chosen in these plots produce minimum dispersion in $\log R_0$ with respect to the 3rd order polynomial fit given by the black dashed lines in panels (c) and (d). Panel (c) corresponds to the sizes inferred from $SB$ whereas sizes from $\Sigma$ are represented in panel (d).  The blue and red lines in panel (d) are used for reference and represent the sequence followed by the galaxies analyzed in \citet{2020arXiv200102689T}. They match seamlessly with the relation that we infer.  
}
\label{fig:getsizes1}
\end{figure*}

The values of $R_0$ inferred from the optimal $SB_0$ and $\Sigma_0$ are  shown in  Fig.~\ref{fig:getsizes1}. It is almost identical to Fig.~\ref{fig:getsizes0} except for using $SB_0=25\,  {\rm mag}\,{\rm arcsec}^{-2}$ (Fig.~\ref{fig:getsizes1}c) and $\Sigma_0=2\,M_\odot$\,pc$^{-2}$ (Fig.~\ref{fig:getsizes1}d). The plots include the 3rd order polynomial fit used to compute the RMS. In addition, Fig.~\ref{fig:getsizes1}d includes the straight lines with the slopes inferred by \citet{2020arXiv200102689T} when using  $\Sigma_0=1\,M_\odot$\,pc$^{-2}$ (the blue and red dashed lines, which correspond to their early and late type galaxies, respectively). They agree extremely well with the slope of the relation that we infer, providing a sanity check for our analysis of the observed NSA data.
Our slope goes from 0.38 at $M_\star\sim 3\times 10^{9}\,M_\odot$ to 0.57 at $2\times 10^{11}\,M_\odot$, which compares extremely well with the slopes inferred by \citet{2020arXiv200102689T}, i.e.,  0.36 and 0.58 respectively.
%

Figures \ref{fig:getsizes0} and \ref{fig:getsizes1} allow us to appreciate the impact  of changing $SB_0$ and $\Sigma_0$ in the recovered relations. The scatter does not change much (RMS varies from 0.057 to 0.054 when $SB_0$ changes from 24 to 25 mag\,arcsec$^{-2}$). The main effect is on the obvious spread among the galaxies with different Sersic indexes, which seems to be smallest at the optimal  $SB_0$ and $\Sigma_0$.

The best values for $SB_0$ and $\Sigma_0$ inferred above are very much consistent with each other. Assuming a constant mass-to-light ratio $M/L$, 
\begin{equation}
  \log\big[\frac{\Sigma}{3.1\, M_\odot\,{\rm pc}^{-2}}\big]=-0.4\,(SB-25)+\log(M/L),
  \label{eq:equivalence}
\end{equation}
with $M/L$ given in solar units and $SB$ measured in the $r$-band in ${\rm mag\, arcsec}^{-2}$ \citep[e.g.,][]{2008ApJ...683L.103B}.  For the optimal $SB_0=24.7$ mag~arcsec$^{-2}$ to be strictly consistent with the optimal $\Sigma_0= 2.4\,M_\odot\,{\rm pc}^{-2}$, Eq.~(\ref{eq:equivalence}) requires  $\log(M/L)\simeq -0.25$. This $M/L$ is right in the middle of the expected ratio for individual galaxies \citep[from -0.4 to +0.6; e.g.,][]{2001ApJ...550..212B}.

\subsection{Galaxy selection bias on the size--mass and size--magnitude relations}\label{sec:selec_bias}

\begin{figure*}
\includegraphics[width=1.0\linewidth]{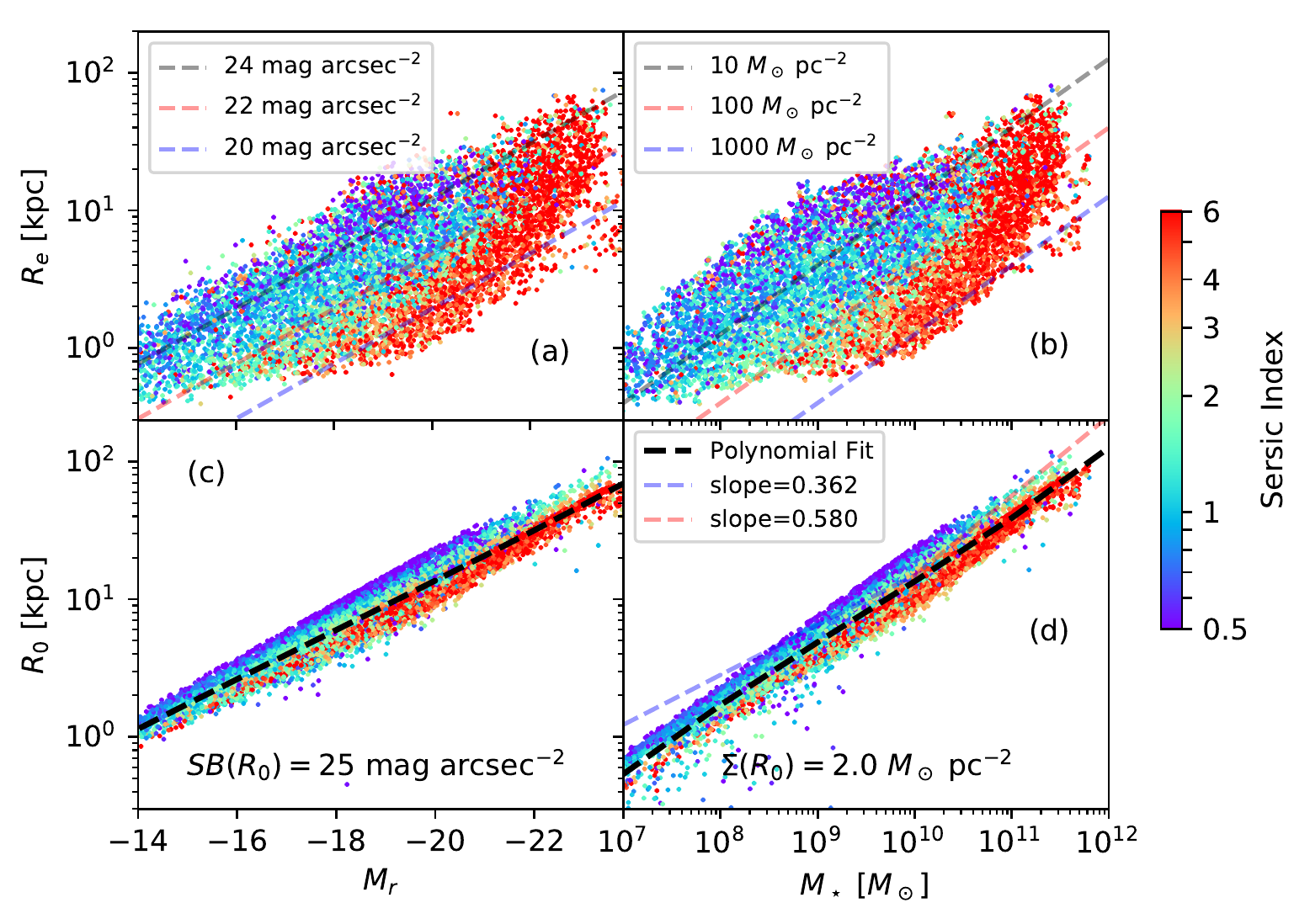}
\caption{
Same as Fig.~\ref{fig:getsizes1} for another sub-sample of the NSA catalog selected so that the galaxies are evenly spread over the observed  $\log R_e$ -- $\log M_r$  plane. $SB_0$ and $\Sigma_0$ in the panels (c) and (d) are the same as in Fig.~\ref{fig:getsizes1}c and \ref{fig:getsizes1}d because the values that produce least scatter are similar for the two different sub-samples. The largest differences with respect to Fig.~\ref{fig:getsizes1} are in the RMS (significantly larger in this case) and on the resulting relations (the black dashed lines). All the straight lines are the same as in Fig.~\ref{fig:getsizes1} and have been included for reference.
}
\label{fig:getsizes3}
\end{figure*}

The galaxies analyzed in the previous sections correspond to a randomly chosen subset of the original NSA catalog. As such, it is biased towards the regions of the parameter space where most of the galaxies reside, and unusual objects are underrepresented. In order to study the impact of this selection criterion on the resulting $R_0$ versus $M_r$ and  $R_0$ versus  $M_\star$ relations, we selected another sample in a completely different way, so that the chosen galaxies are evenly spread in the  $\log R_e$ -- $\log M_r$  plane. The selection was made dividing the  $\log R_e$ -- $\log M_r$ plane in $50\times 50$ equal pixels. We include all the galaxies in those pixels having between 3 and 10 objects, and 10 randomly selected galaxies in those pixels having more than 10 objects. These bin size and threshold were tuned to end up with a sample having a number of objects comparable to the random selection described in the previous sections; of the order of $~$7900 galaxies. Figure~\ref{fig:getsizes3}c shows the scatter plot $R_e$ versus $M_\star$ of the objects thus selected and, despite the fact it contains a similar number of galaxies as in Fig.~\ref{fig:getsizes1}c, the points are far more spread out, covering the whole range of observed parameters.  (Figures~\ref{fig:getsizes4}, top and bottom, visually capture the spread of the two different samples that we are considering.)

The analysis carried out with the previous sample was repeated to find out the values for  $SB_0$ and $\Sigma_0$ that minimize the scatter of the $R_0$ versus $M_r$ and  $R_0$ versus  $M_\star$ relations. The optimal values turn out to be very similar to those inferred in Sect.~\ref{sec:least} for the main NSA sub-sample studied in the paper.  In this case, we obtain $SB_0=24.6\pm 0.5$ mag\,arcsec$^{-2}$ and $\Sigma_0=1.3^{+1.2}_{-0.6}\,M_\odot\,{\rm pc}^{-2}$. The main differences are (1) in the RMS, much larger in this case (of the order of 0.10, to be compared with 0.06), and (2) in the resulting relations (cf. the black dashed lines in Figs.~\ref{fig:getsizes1}c and \ref{fig:getsizes1}d with Figs.~\ref{fig:getsizes3}c and \ref{fig:getsizes3}d). In other words, both the RMS and the relation seem to depend on the used galaxy sample.

The question arises as to whether the sample portrayed in Fig.~\ref{fig:getsizes3} is representative of the local galaxies. Probably it is not since the selection was strongly biases towards including outliers in the $R_e$ versus $M_\star$ relation. This explains the large difference between the relations $\log R_0=\log R_0(\log M_\star)$ found with the two subsamples analyzed in this paper as compared to the agreement between Fig.~\ref{fig:getsizes1}d and the sample of galaxies used by \citet{2020arXiv200102689T}. Keeping in mind the difference in galaxy sample and method of analysis, the agreement also indicates that the resulting relations is probably more {\em universal} than what one may naively infer from the differences existing between Figs.~\ref{fig:getsizes1}d and \ref{fig:getsizes3}d.

 For the sake of comprehensiveness, we also want to report on another sanity check carried out to understand the selection bias. The azimuthal averages used for the NSA Sersic fits are corrected for galaxy inclination  (Sect.~\ref{sec:observations}). This is equivalent to working with face-on galaxies, as we do along the manuscript. However, the correction assumes axisymmetric optically-thin galaxies, which  in some cases may not be a good approximation. In order to study the impact of the inclination correction, we selected from the NSA full sample two additional subsets  biased to extreme galaxy inclinations. One having $b/a > 0.8$ and another one with $0.2 < b/a < 0.4$, where $b/a$ stands for the axial ratio of the outer isophotes as provided by NSA. No major differences are observed except for a slight systematic decrease of the Sersic index for the highly inclined galaxies ($0.2 < b/a < 0.4$). This change has an impact on the optimal  $\Sigma_0$ and $SB_0$, but it is much smaller than the increase of spread displayed in Fig.~\ref{fig:getsizes3}.

%
%
\section{Can we use galaxy size as a proxy for galaxy mass?}\label{sec:proxy_mass}

\begin{figure*}
  \centering
  \includegraphics[width=0.8\linewidth]{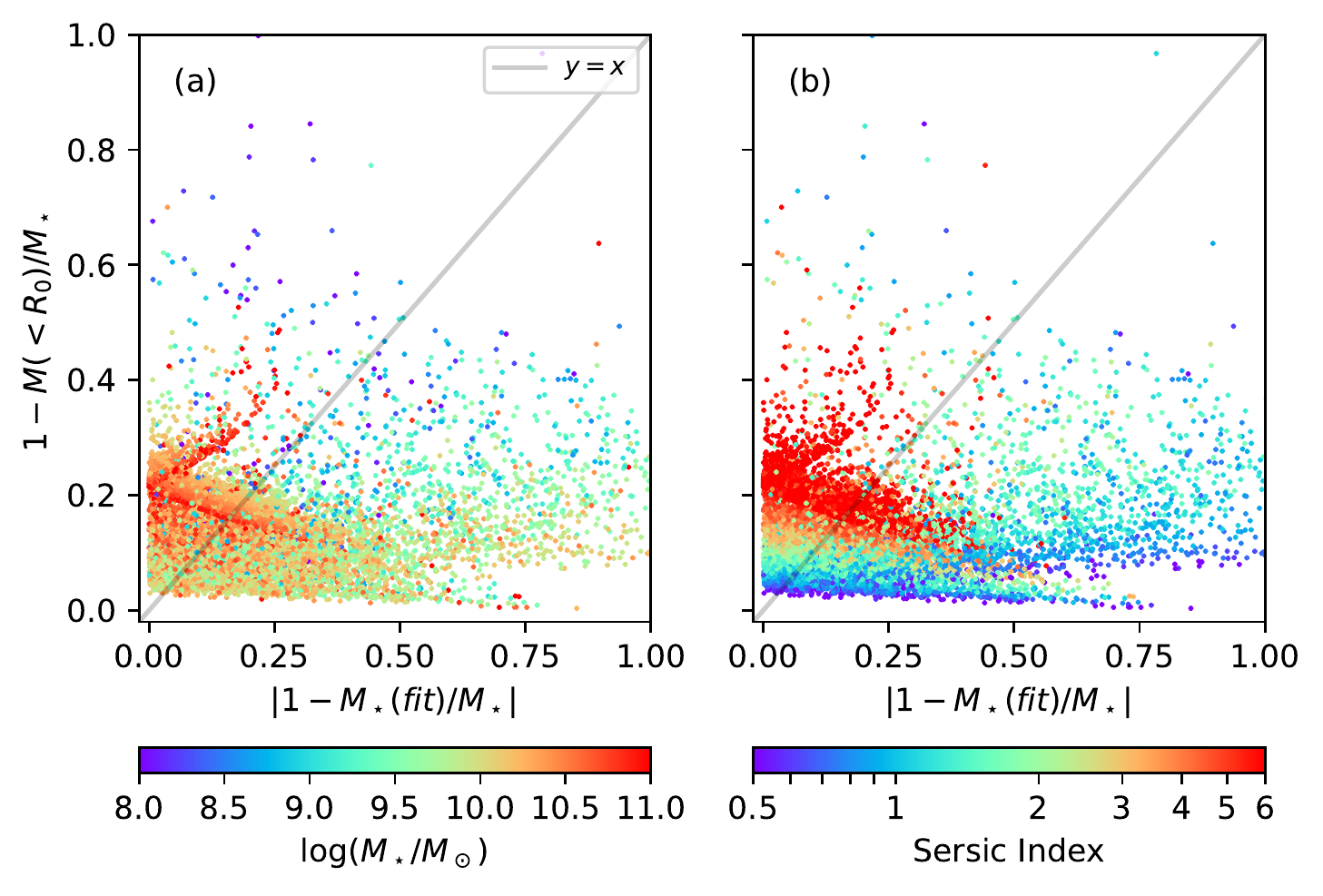}
\caption{(a) Error in the mass inferred from photometry versus error in the empirical calibration $M_\star=M_\star(R_0)$ when both approaches use the same images.  The points are color coded according to $M_\star$. The galaxies are the same as in Fig.~\ref{fig:getsizes1}, and the used empirical calibration is the 3rd order polynomial included in Fig.~\ref{fig:getsizes1}d. The solid line represents the one-to-one relation, and around 40\,\% of all galaxies lie above this line. (b) Same as (a) color coded with the Sersic index of the galaxy.}
  \label{fig:getsizes_fit}
\end{figure*}

Since $R_0$ is a tight function of $M_\star$, the inverse of this relation could  be calibrated to infer $M_\star$ from $R_0$. A possible drawback of this approach is the need for deep images to determine $R_0$.  One may think that getting them is more time-consuming that the direct determination of $M_\star$ from photometry. However, the depth of the images to get a good photometric mass estimate and to get $R_0$ is similar, therefore, from the point of view of the observing time demand, both approaches are comparable. This can be shown assuming that the images used to determine $M_\star = M_\star(R_0)$ and $M_\star$ from photometry are the same, and then working out the error in $M_\star$ in both cases.   The arguments goes as follow:  reaching the depth needed to determine $R_0$ provides photometric information on the galaxy only for $R < R_0$, therefore, assuming $L/M$ to be known, the stellar mass inferred from photometry is 
\begin{equation}
  M(< R_0)=2\pi \int_0^{R_0}\,R\,\Sigma(R)\,dR,
\end{equation}
yielding a relative error of
\begin{equation}
  1-M(< R_0)/M_\star = 1-\gamma(2n,\xi)/\Gamma(2n),
  \label{eq:errors}
\end{equation}
with $\xi=b\,(R_0/R_e)^{1/n}$. The derivation of  $M(< R_0)/M_\star$ in terms of gamma functions when $\Sigma$ follows a Sersic profile can be found elsewhere \citep[e.g.,][]{2001MNRAS.326..869T,2005PASA...22..118G}.
Figure~\ref{fig:getsizes_fit} compares the relative error in Eq.~(\ref{eq:errors}) with the error of the empirical calibration represented by the 3rd order polynomial in Fig.~\ref{fig:getsizes1}d (the black dashed line). The scatter plot in Fig.~\ref{fig:getsizes_fit} contains all the galaxies in the NSA reference sample (Fig.~\ref{fig:getsizes1}). About 40\,\% of the galaxies lie above the diagonal line in Fig.~\ref{fig:getsizes_fit}, meaning that their photometric mass error is larger than the error in their mass inferred from the empirical calibration. Despite naive, this simulation tells us that using $R_0$ as a proxy for $M_\star$ often is as good or better than estimating $M_\star$ from photometry.  Moreover, the empirical calibration has potential for improvement  when combined with morphological classification. The dispersion in $M_\star=M_\star(R_0)$ to be achieved when using only early types ($n\sim 1$) or only late types ($n\sim 4$) can potentially be much smaller than the one portrayed in the abscissae of Fig.~\ref{fig:getsizes_fit}. In addition, the availability of images in various bands can provide independent estimates of $M_\star$, improving the overall accuracy. On the contrary, the error Eq.~(\ref{eq:errors}) is only a lower limit to the true error since it neglects uncertainties in $M/L$ as well as additional photometric errors associated with the reduction of the images. For example, the presence of bulges and bars complicate the interpretation of the photometry whereas it does not affect the galaxy outskirts from which $R_0$ is derived.  
Thus, getting $M_\star$ directly from photometry or through $R_0$ require images with similar depth. Choosing one way or another is not an issue of improving efficiency but a question of practice and convenience. The use $R_0$ as a proxy for $M_\star$ may be particularly useful at high redshift, however, the empirical calibration to be employed remains to be worked out. The usefulness of  local calibrations for high redshift galaxies has not been proven yet.
  Moreover, the observations of high redshift galaxies have limitations associated with the decrease of angular resolution and surface brightness, which could burden the technique to the point of making it unfeasible. A way to evaluate the potential problems and to carry out the required calibration would be repeating the analysis in the present paper using high-redshift databases
\citep[e.g., CANDELS; ][]{2011ApJS..197...35G}, a work that we are currently undertaking.

%
\section{Discussion and conclusions}\label{sec:discussion}

Looking for a physically motivated definition of galaxy size, \citet{2020arXiv200102689T} found that the size--mass relation reduces the scatter considerably when, rather than using $R_e$, the size at a given surface density is used. They employ $\Sigma_0 = 1 M_\odot {\rm pc}^{-2}$ arguing that it is set by the threshold needed for the gas to form stars (see Sect.~\ref{sec:intro}).  Answering the question of why the scatter drops and why it happens at this $\Sigma_0$ is the motivation of our work.   

We split the question, and so the answer, in three different parts: (1) Why does the scatter drop?, (2) why does the scatter drop so much?, and (3) why is there an optimal $\Sigma_0$ producing least scatter? 
(1) A reduction in the scatter is to be expected because the mass surface density profiles are continuous monotonically decreasing functions of radius, and so, any two functions containing the same $M_\star$ have to cross at a particular radius. Choosing $\Sigma_0$ close to the crossing point necessarily reduces the scatter (Sect.~\ref{sec:theprinciple}).
(2) Why does the scatter drop so much? Because the mass surface density profiles  are well represented by Sersic profiles, combined with the anti-correlation between $R_e$ and $n$ in the observed galaxies (Sect.~\ref{sec:math}). In a way this answer just changes the question from {\em why does the scatter drop so much?} to {\em why do mass density profiles follow Sersic functions?}, a question which does not seem to have a simple answer either \citep[e.g,][]{2014ApJ...790L..24C,2017IAUS..321...87N}.
(3) Why is there an optimal $\Sigma_0$? The short answer is, again, because the observed spread and anti-correlation in $R_e$ and $n$ conspire to provide a suit spot at  $\Sigma_0$. However, this answer has to be expanded to provide the full view. As judged from the NSA catalog of galaxies, which contains a comprehensive representation of the local Universe, $\Sigma_0$ may have a significant range of values and still provide similarly low scatter ($2.4_{-0.9}^{+1.3}\,M_\odot\,{\rm pc}^{-2}$; Sect.~\ref{sec:least} and Fig.~\ref{fig:optimum}). Moreover, the scatter and the relation depend quite significantly on the galaxy set (Sect.~\ref{sec:selec_bias}), and so on the Sersic indexes of the chosen galaxies. The scatter tends to be smaller for larger Sersic indexes, as predicted from theory (Sect.~\ref{sec:math}).  Thus, from the analysis carried in Sect.~\ref{sec:observations}, different galaxy types and galaxy sets seem to have moderately different values of $\Sigma_0$. Note that this fact and our optimal value for $\Sigma_0$ are not in contradiction with the value and the interpretation for $\Sigma_0$ given by \citet{2020arXiv200102689T}.{
  The masses used in this work were derived under the assumption of a {\em diet} Salpeter IMF, which provides masses some 1.4 times systematically larger than the Chabrier IMF employed by \citet{2020arXiv200102689T} \citep[see][]{2003ApJS..149..289B}. Thus, their  $1\,M_\odot\,{\rm pc}^{-2}$ is equivalent to  $1.4\,M_\odot\,{\rm pc}^{-2}$ in our scale, which fits in the range of values that we find.
In addition, galaxies with different star-formation histories and environments are expected to respond differently to the existence of a gas density threshold, naturally leading to the observed spread.

Since galaxy surface brightness profiles also follow Sersic functions, all the above conclusions apply to surface brightness profiles as well. The scatter of the size--absolute magnitude relation also gets reduced adopting a size definition based on a constant $SB_0$. We find the scatter to be smallest for  $SB_0= 24.7\pm 0.5\,{\rm mag\,arcsec^{-2}}$ in the $r$-band, which is very much consistent with the optimal $\Sigma_0$ given a reasonable $M/L$ (Sect.~\ref{sec:least}). Actually, the scatter of the $R_0$ versus $M_r$ relation inferred from $SB$ (0.054\,dex) is slightly smaller than the scatter  of the  $R_0$ versus $M_\star$ relation inferred from $\Sigma$ (0.056 \,dex).

We use galaxies in the NSA catalog to test the theoretical ideas worked out in Sects. ~\ref{sec:theprinciple} and \ref{sec:math}. For example, the prediction that the slope of the size--mass relation increases with increasing Sersic index (Eq.~[\ref{eq:slope}]). They are also employed to show that $R_e$ and $n$ are anti-correlated as assumed to predict the drop of scatter modeled in Fig.~\ref{fig:intuition5b}c. However, the argument that the scatter in the size--mass relation is reduced in NSA galaxies has some degree of circularity.   The galaxy catalog does not provide the full $\Sigma$ profile, therefore, in order to derive $R_0$ from $\Sigma$, we have to assume that the profiles follow a Sersic function. Thus, the fact that the NSA galaxies get their scatter reduced when employing $R_0$ rather than $R_e$ may be artificial since it relies on the assumption that the galaxies follow Sersic functions down to $\Sigma_0$. This circularity in the argumentation is broken because the function $R_0=R_0(M_\star)$ that we infer (Fig.~\ref{fig:getsizes1}d) is very much consistent with the relation worked out by \citet{2020arXiv200102689T}, who do not make any assumption on $\Sigma$ to get $R_0$. Our relation agrees with theirs on the slope as well as on the magnitude of the scatter. The best RMS that we derive is close to the intrinsic scatter of the relation that they estimate, of the order of 0.06\,dex.
In this sense,  we note that a comparison of the observed relations with cosmological numerical simulations \citep{2015MNRAS.446..521S} is underway, with a very good overall agreement. The model galaxies quantitatively reproduce the observed trends within errors (Dalla Vecchia 2020, private communication).

As a by-product of this work, and considering the narrowness of the size--mass relation, we propose to use $R_0$  to measure $M_\star$. This alternative approach  may be useful in some cases, e.g., when multi-band observations are not available and $M/L$ cannot be easily estimated. In Sect.~\ref{sec:proxy_mass}, we show that getting $M_\star$ directly from photometry or through $R_0$ requires images with similar depth. Thus, choosing one way or another is mostly an issue of practice and convenience. Using $R_0$ as a proxy for $M_\star$ may be particularly useful at high redshift, however, the required empirical calibration of the relation remains to be worked out.

%
%

%

\section*{Acknowledgements}
%
%
I thank Ignacio Trujillo for illuminating discussions on the topic of the paper, and for comments on the original manuscript.
Thanks are also due to Claudio Dalla Vecchia for allowing me to mention work in progress, and to an anonymous referee for suggestions to improve various passages of the manuscript.
The work has been partly funded by the Spanish Ministry of Economy and Competitiveness (MINECO), project AYA2016-79724-C4-2-P (ESTALLIDOS). 
I acknowledge extensive use of the NASA Sloan Atlas$^{\ref{foot:1}}$ (V1\_0\_1) as well as the {\em astrophy} tool suite \citep{2013A&A...558A..33A}.
%







\appendix
\section{Equivalence of mass surface density and surface brightness profiles}\label{app:appa}

The light surface density profile, $\Sigma_L$, follows a Sersic function formally identical to the mass surface density profile $\Sigma$ (Eqs.~[\ref{eq:sigma}] and [\ref{eq:sersic}]) replacing $M_\star$ with the absolute luminosity $L$. $\Sigma_L$ is independent of the distance to the galaxy $d$, and it is also proportional to the energy flux per unit solid angle received on earth, therefore, except for a multiplicative constant, it provides the surface brightness, i.e.,
\begin{equation}
  SB(s) = -2.5\log(\Sigma_L)+c = SB_e+\big[(s/r_e)^{1/n}-1\big]\,2.5\,b\log e,
\end{equation}
with $s$ and $r_e$ representing the apparent distance from the center of the galaxy and the apparent effective radius, respectively ($s=R/d$ and $r_e=R_e/d$). The symbol $c$ is a calibration constant, and  
\begin{equation}
 SB_e=SB(r_e)=M_r+2.5\,\log\big[2\pi\,n\,R_e^2\,e^b\,\Gamma(2n)/b^{2n}\big].
\end{equation}
As usual, $M_r$ stands for the absolute magnitude corresponding to $L$,
\begin{equation}
 M_r=-2.5\,\log L + c.
\end{equation}
Defining the apparent size of the galaxy $r_0$ as the apparent radius at a constant surface density $SB_0$, 
\begin{equation}
 SB(r_0) = SB_0,
 \end{equation}
 then the physical size $R_0$ can be retrieved from known parameters ($M_r$, $R_e$, and $n$) as,
 \begin{equation}
   \log(R_0/R_e)= n\log\big(1+\frac{SB_0-SB_e}{2.5\,b\log e}\big),
 \end{equation}
 keeping in mind that the ratio of apparent sizes $r_0/r_e$  is identical to the ratio of true sizes $R_0/R_e$. 


\bsp	
\label{lastpage}
\end{document}